\documentclass[fleqn,10pt]{wlscirep}
\usepackage[utf8]{inputenc}
\DeclareUnicodeCharacter{0301}{\'{e}}
\usepackage[T1]{fontenc}
\usepackage[draft]{fixme}
\usepackage{booktabs}				
\usepackage{dcolumn}				
\usepackage{xcolor}
\usepackage{soul}
\usepackage{enumerate}

\usepackage[style=nature]{biblatex}  
\usepackage{fancyhdr}
\usepackage{lastpage}
\usepackage{comment}
\excludecomment{mycomment}

\newcommand{\beginsupplement}{%
        \setcounter{table}{0}
        \renewcommand{\thetable}{S\arabic{table}}%
        \setcounter{figure}{0}
        \renewcommand{\thefigure}{S\arabic{figure}}%
     }

\title{Spontaneous emergence of groups and signaling diversity in dynamic networks}

\author[1]{Zachary Fulker}             
\author[2]{Patrick Forber}             
\author[3]{Rory Smead}                 
\author[1,*]{Christoph Riedl}          
\affil[1]{Network Science Institute, Northeastern University, Boston, MA}
\affil[2]{Tufts University, Medford, MA}
\affil[3]{Northeastern University, Boston, MA}
\affil[*]{c.riedl@neu.edu}

\begin{abstract}

We study the coevolution of network structure and signaling behavior. We model agents who can preferentially associate with others in a dynamic network while they also learn to play a simple sender-receiver game. We have four major findings. First, signaling interactions in dynamic networks are sufficient to cause the endogenous formation of distinct signaling groups, even in an initially homogeneous population. Second, dynamic networks allow the emergence of novel {\em hybrid} signaling groups that do not converge on a single common signaling system but are instead composed of different yet complementary signaling strategies. We show that the presence of these hybrid groups promotes stable diversity in signaling among other groups in the population. Third, we find important distinctions in information processing capacity of different groups: hybrid groups diffuse information more quickly initially but at the cost of taking longer to reach all group members. Fourth, our findings pertain to all common interest signaling games, are robust across many parameters, and mitigate known problems of inefficient communication. 

\end{abstract}

\begin{document}

\flushbottom
\maketitle
\thispagestyle{empty}


Signaling is ubiquitous in the living world. Individual molecules trigger complex metabolic cascades or immune responses. Animals routinely use signals to alert their group to the presence of predators, to mediate interactions between parents and offspring, and attract mates \cite{jms-harper2003,searcy2005,bradbury2011}. Humans have harnessed the power of signaling to an unparalleled degree, giving rise to cultural evolution and evolving systematic languages that enable a broad range of expression \cite{lieberman2007quantifying,Pinker1990,pinker2003language,nowak1999evolution}. When communication involves common interest, simple sender-receiver games show how reliable signaling can emerge and evolve to transmit information effectively. Although signaling games are highly idealized, the results are so robust that they are expected to generalize to almost all cases of common interest signaling, from molecular cell-to-cell signaling to human language. This makes them a powerful model with broad implications. While these games have been analyzed extensively in randomly mixing populations and on static network structure \cite{lewis1969convention,skyrms2010signals}, past work has not investigated sender-receiver games on dynamic networks. Furthermore, past work has not explored the evolutionary viability of hybrid signaling strategies.

We explore the evolution of signaling in contexts where agents can choose their communication partners. We aim to better understand how interactions between agents create order out of an initially disordered situation \cite{Castellano2009}. An ongoing area of research in this domain focuses on signaling and communication dynamics: how do distinct (social) groups and ways of communicating within them emerge? To answer this we combine models of signaling games \cite{skyrms2010signals,skyrms2014evolution} with dynamic networks \cite{skyrms2000dynamic} to investigate the coevolution between strategies and network structure. Structural change is a common feature of the real world. Social organisms often can choose their interaction partners and the frequency with which they interact \cite{chaudhuri2009experiments}. This allows the emergence of structured interactions resulting from the adaptive nature of agents \cite{perc2010coevolutionary,vespignani2012modelling,Fulker2021,foley2021avoiding,tomassini2010,zimmermann2004coevolution}. We show that signaling interactions in dynamic networks are sufficient for the endogenous formation of distinct groups from initially homogeneous populations. We also see novel signaling behavior: groups emerge using combinations of strategies that would fail to communicate in standard settings. Our results pertain to all common interest signaling games, are robust across many parameters, and mitigate known problems of inefficient communication. The model suggests a new possible explanation for the presence of signaling diversity and for the emergence of structure from initially unstructured states.

\section*{The Model}

\begin{figure}[t!] \centering
    \includegraphics[width=.9\linewidth]{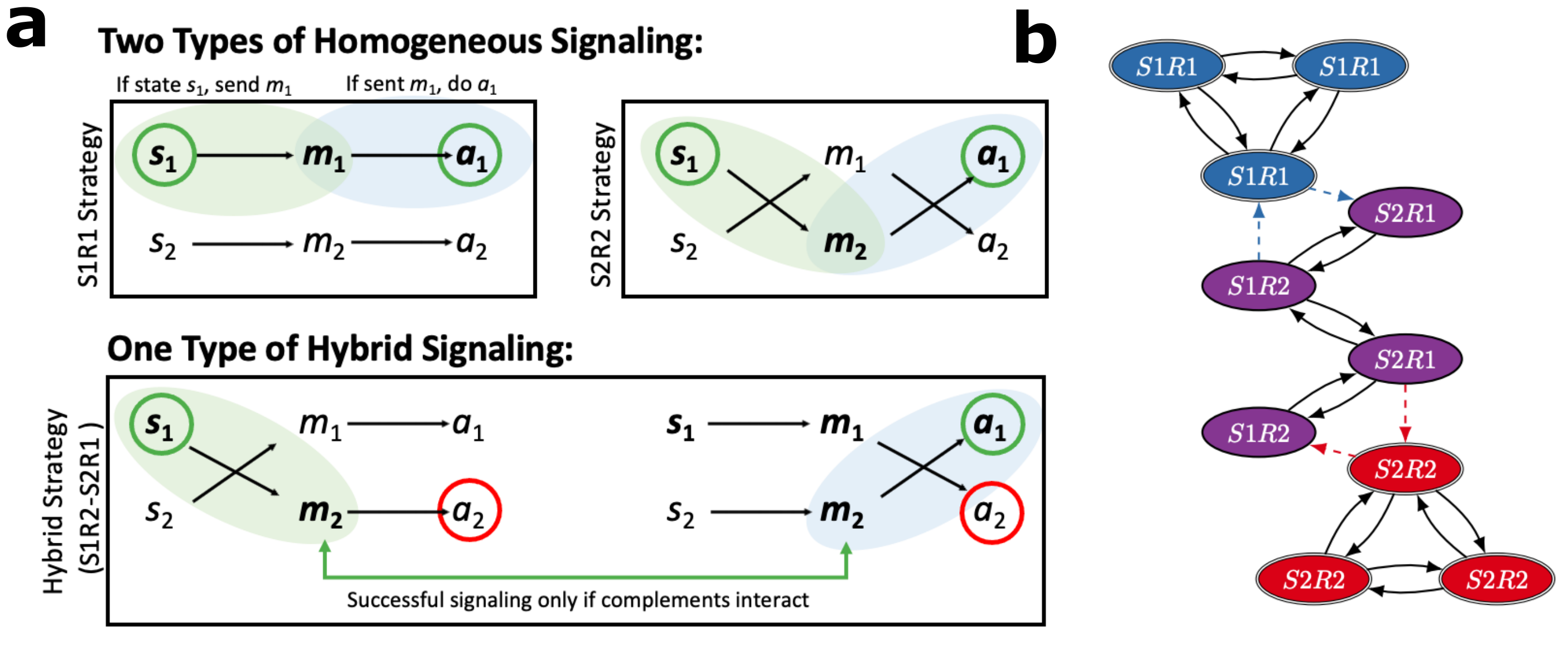}
    \caption{\textbf{Different strategies to achieve signal coordination. a)} \textit{Homogeneous} signaling (top): complement strategies match. \textit{Hybrid} signaling: complement strategies differ. Agents using hybrid strategies can coordinate successfully with complementary types, but they take the wrong action if when interacting with themselves  (red circles). 
    \textbf{b)} Homogeneous agents interact optimally with any agent using the same strategy, while hybrid agents  only interact optimally with the complementary types (solid black arrows). Interactions between the hybrid types and the homogeneous types receive non-zero payoffs, but are nearly eliminated by the end of the simulation (dashed arrows).} 
    \label{fig:panel0}
\end{figure}

Suppose two agents (e.g., individuals or cells) have a common interest in communicating about the state of the world (Fig.~\ref{fig:panel0}). The sender observes the state of the world and sends a signal to the receiver. The receiver observes the signal but not the state and must select an action. There is a single correct action for each possible state and the signals have no natural salience. The simplest version of the game involves two equiprobable states ($s_1, s_2$), two possible messages ($m_1, m_2$), and two actions ($a_1, a_2$). This is a $2\times2\times2$ Lewis sender-receiver game \cite{lewis1969convention}.  There are four possible sender strategies the agents might employ. They may: send $m_1$ in $s_1$ and $m_2$ in $s_2$ (S1), $m_2$ in $s_1$ and $m_1$ in $s_2$ (S2), $m_1$ regardless of the state (S3), or $m_2$ regardless of state (S4). The first two are \textit{separating} strategies that use the signals to discriminate between the states, the second two are \textit{pooling} strategies. The receiver also has four  strategies: $a_1$ when receiving $m_1$ and $a_2$ when receiving $m_2$ (R1), $a_2$ when receiving $m_1$ and $a_1$ when receiving $m_2$ (R2), $a_1$ regardless (R3), or $a_2$ regardless (R4). These are the \emph{separating} and  \emph{pooling} strategies, respectively. Actions are successful if $a_1$ is done in $s_1$ and $a_2$ in $s_2$. Therefore, success is only guaranteed when both the sender and receiver use complementary separating strategies. A pair of appropriately complementary strategies is known as a \textit{signaling system} (shown in top row of Fig.~\ref{fig:panel0}a). Agents in our model have a strategy for both sender (S) and receiver (R) roles. In the $2\times2\times2$ game, this results in 16 total strategies (four for each role) and we label these strategies by each of their components (e.g., S1R3).

Traditional evolutionary models with large, randomly mixing populations invariably yield one of the two {\em homogeneous} signaling systems (S1R1 or S2R2; Fig.~\ref{fig:panel0}a) \cite{hofbauer2008feasibility,skyrms2014evolution,skyrms2010signals}. We model a finite population of agents that choose their own interaction partners and update both their signaling strategies and their network through reinforcement learning \cite{skyrms2000dynamic,skyrms2004stag,FoleyEtAl2018}. Reinforcement learning has been identified as a biologically plausible mechanism for iterative decision-making settings \cite{erev1995, erev1998predicting, Gershman2017}.

When Lewis sender-receiver games are played on dynamic networks, we observe signaling systems that are not seen in unstructured populations. Agents may rely on network connections to solve communication problems rather than converging on a shared signaling system. This allows for the evolution of a {\em hybrid} solution Fig.~\ref{fig:panel0}a). This requires two different but complementary signaling strategies to be paired via network connections. For example, an agent may send $m_1$ in state $s_1$, but as receiver responds to $m_1$ with $a_2$ (strategy S1R2). Such an agent would do poorly when paired with a similar agent, but communicates successfully if paired with one using an inverted strategy (S2R1). These hybrid solutions do not emerge in large randomly mixing populations because they cannot effectively communicate with others using the same strategy. In dynamic networks, however, agents of one type can reliably pair with agents of the other type and vice versa while avoiding agents of the same type (Fig.~\ref{fig:panel0}b).

\begin{figure}[t!] \centering
    \includegraphics[width=.9\linewidth]{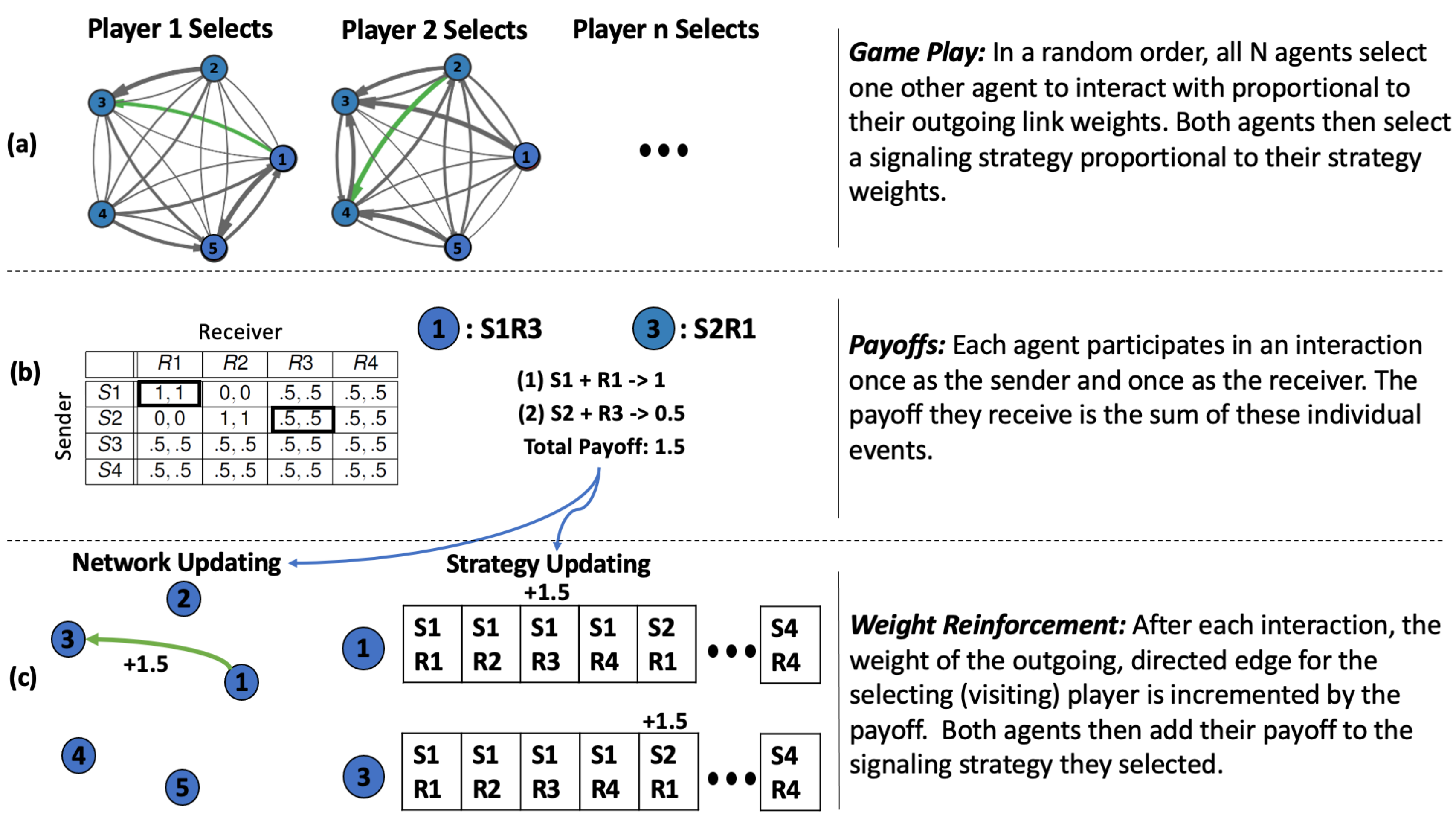}
    \caption{
        \textbf{Game play and updating mechanism.} In each round agents must 
        \textbf{(1)} select an interaction partner, 
        \textbf{(2)} earn a payoff based on the expected outcome, and 
        \textbf{(3)} update their network and strategy weights based on the payoff.
        }
    \label{fig:panel1}
\end{figure}

Groups using hybrid signaling systems are stable in our model. To see this, suppose that there is a group of agents such that each almost always plays S1R2 or almost always S2R1. Suppose further that S1R2 types almost exclusively interact with S2R1 types and vice versa. In this case, each type is then receiving a near maximum expected payoff in each signaling interaction. The strategy combination represents a strict Nash equilibrium where neither type would prefer to switch their strategy given their partners strategies as any other strategy would do strictly worse. Moreover, S1R2 would be indifferent between switching among S2R1 partners, but would do strictly worse with a non-S2R1 partner (and similarly for S2R1 pairing with S1R2). Since reinforcement learning will never tend toward strictly worse payoffs \cite{erev1995,beggs2005convergence}, we should expect both types to continue visiting the complementary type almost exclusively and to not shift strategies. Similar reasoning, though with symmetric rather than complementary strategies, applies to homogeneous groups as well. Our numeric analyses show that both hybrid and homogeneous groups emerge reliably and that hybrid groups contribute to more intricate networks with more diversity of signaling behavior.

The stability of multiple groups with different signaling strategies is straightforward to demonstrate theoretically. Suppose we have a population with three exclusive groups arranged (similar to that in Fig.~\ref{fig:panel0}b) with two opposite homogeneous groups and one hybrid. Suppose also that each agent only visits others with complementary strategies for optimal payoff. Given this arrangement, all agents would do strictly worse by adopting any other signaling strategy. No alternative network connections generate an advantage. The only alternative connections that do equally well are variations among other agents with the same strategies as the current partners. Consequently, only networks that preserve current group composition are viable alternatives. Therefore, coexisting homogeneous and hybrid signaling system arrangements among multiple groups are stable. We investigate whether and with what frequency these arrangements can evolve in populations using numerical simulations.


\section*{Methods}

\paragraph{Setup and game play.} 
We model a set of $N$ agents that, over $10^6$ rounds, play sender-receiver games with each other. Each round, every agent independently chooses one interaction partner to visit (Fig.~\ref{fig:panel1}). Agents may receive multiple visitors (between 0 and $N-1$) and initiate one visit themselves. Within interactions agents select strategies and receive payoffs. Both agents earn the expected payoff given the selected strategies assuming equal chance of being the sender or receiver in the game. For most simulations we also assume equiprobable states of the world and we note the cases where this assumption is relaxed. After each round each agent's network connections and strategy choices are updated based on the payoffs received. 

Each agent $i$ has a vector of strategy weights $(s_{1}, s_{2},..., s_{16})$ where $s_n$ represents one of the $16$ possible sender-receiver strategy pairs ($S1R1, S1R2, ..., S4R4$). Each agent also  has a vector representing their outgoing link weights used to choose which player to visit: $(w_{i1}, w_{i2},...,w_{in})$ where $w_{ij}$ represents the weight related to agent $i$ visiting agent $j$. Self-visits are not allowed $(w_{ii} = 0)$. Initial strategy weights are set uniformly and held constant across all simulations $(S=5)$. Initial network weights are also uniform and constant across all simulations $(L=0.19)$. 

\paragraph{Updating.}

Both the strategy weights and network weights are updated via Roth-Erev reinforcement learning \cite{erev1998predicting}.  Partners are chosen with a probability proportional to past payoffs when visiting that partner. Likewise, strategies are chosen with a probability proportional to past payoffs when using that strategy.

At the end of a round an agent's strategy weight $s_i$ is increased by adding any payoffs received when using $s_i$ across all interactions. If a strategy was not used or zero payoff was received then no weight is added. If multiple interactions occurred, all strategies used are updated accordingly.

An agent's outgoing network weights are updated based on the interaction they initiated. Payoffs from that interaction are added to the network weight corresponding to the partner they chose. Interactions initiated by others are not used for updating. This asymmetry represents a situation where individuals choose who to approach, but not who approaches them. Note that the weighted vector represents network links as a matter of degree, in contrast to dynamic network models that represent links as discrete and the dynamics describe patterns of link breakage and formation \cite{Goyal2005, Pacheco2006, Rand2011}. Our approach allows for more realistic continuous network ties \cite{Granovetter1973, Barrat2004, Yook2001} and does not require any limit on number of network connections which can affect results \cite{Rand2014}.

Additional parameters affect updating by reinforcement: discounting ($\delta$), error ($\epsilon$), and learning speed multipliers ($NLS$ and $SLS$). Discounting reduces past learning weights as more reinforcement occurs, gradually allowing agents to forget old preferences. Errors (mistakes, noise) occur when an agent selects a partner or strategy at random rather than according to their weights. Both impact the stability and long-run behavior of reinforcement learning \cite{erev1995, erev1998predicting, Gershman2017, skyrms2000dynamic}. Finally, learning speed multipliers allow us to explore relative differences in strategy and network updating speed.

\paragraph{Formal updating rules.} When selecting an interaction partner for a given round, the probability of choosing agent $j$ is proportional to the current network weights:
\begin{equation}
    Pr(j) = (1 - \epsilon) \frac{w_{ij}}{\sum_{k}w_{ik}} + \epsilon\frac{1}{|N-1|}
\end{equation}
where $\epsilon$ is the error rate, $N$ is the set of agents, $j \in N$ and $k \in N$. 

When selecting a strategy for a given round, the probability of choosing strategy $s_{n}$ is likewise proportional to the current strategy weights:
\begin{equation}
    Pr(n) = (1 - \epsilon) \frac{s_{n}}{\sum_{k}s_{k}} + \epsilon\frac{1}{|S|}
\end{equation}
where $S$ is the set of signaling strategies and $k \in S$.

After each round, link weights are updated by discounting the prior weight by a factor ($\delta$) and adding any received payoff ($\pi$):
\begin{equation}
    w_{ij}' = (1-\delta)w_{ij} + NLS\pi_{ij}.
\end{equation}
Here, $w_{ij}'$ is the link weight after updating and $\pi_{ij}$ is the payoff for agent $i$ visiting agent $j$ ($\pi_{ij}=0$ if link was not selected). $NLS$ is the rate of network learning, set to $NLS=1$ by default, higher values result in faster learning (this is equivalent to increasing payoffs). All link weights are updated simultaneously at the end of each round.

Strategy weights are also updated by discounting the prior weight by a factor ($\delta$) and adding the received payoff ($\pi$):
\begin{equation}
    s_{n}' = (1-\delta)s_{n} + SLS\pi_{n}.
\end{equation}
Here, $s_{n}'$ is the strategy weight after updating and $\pi_{n}$ is the sum of all payoffs an agent earned while using strategy $s_{n}$ in a given round. $SLS$ is the rate of strategy learning, set to $SLS=1$ by default. Just as for link weights, all strategy weights are updated simultaneously at the end of a round, reflecting the outcome of every interaction an agent was a part of during that round.

\section*{Results}

\begin{figure}[b!] \centering
    \includegraphics[width=.8\linewidth]{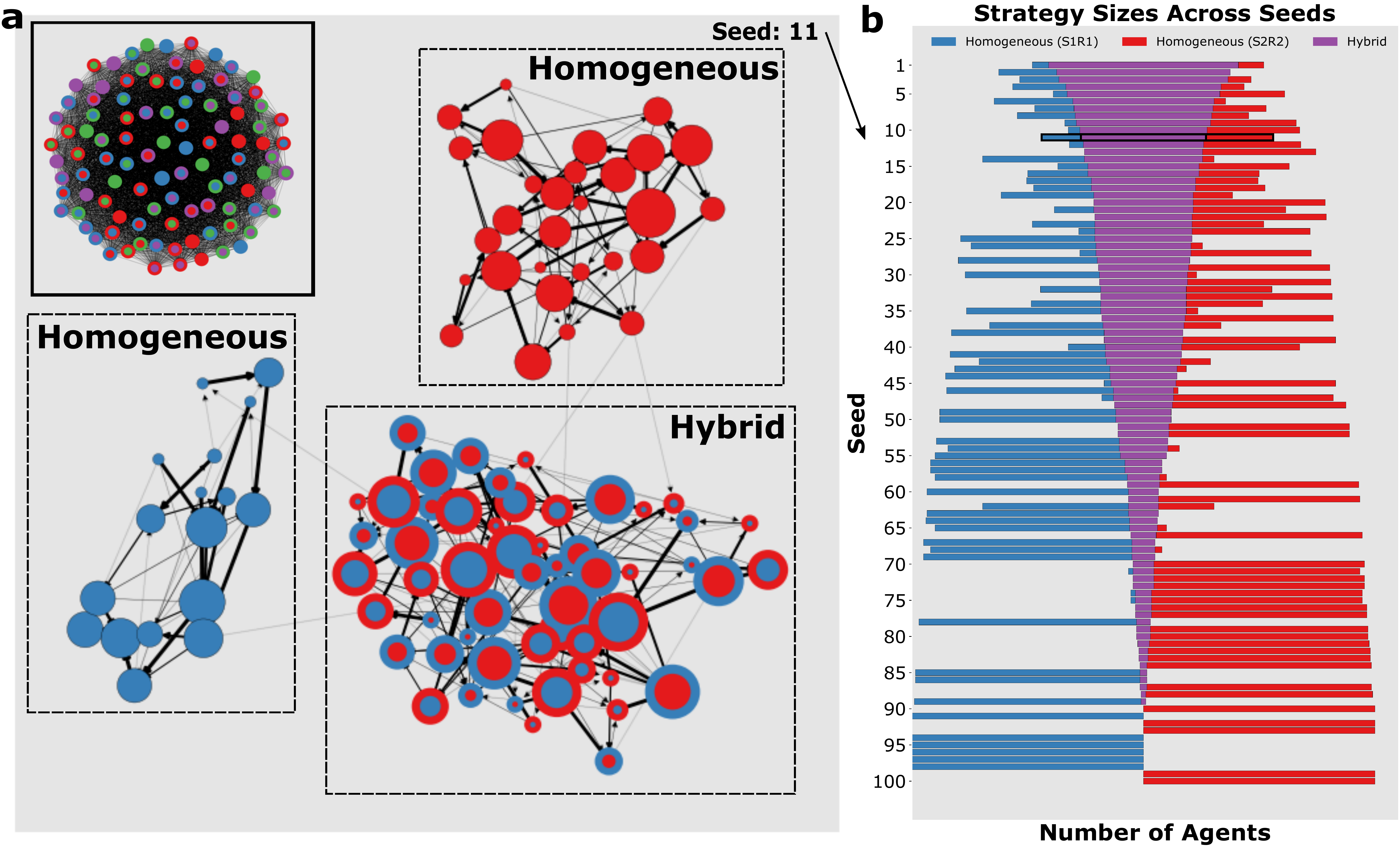}
    \caption{
        \textbf{Emergence of Signaling Groups.} \textbf{a)} Example outcome from one simulation of the $2\times2\times2$ game using force-directed layout. Inner and outer node colors represent an agents primary sender and receiver strategy. \textit{Inset.} Population starts from uniform network ties and strategy weights. \textit{Main.} Population separates into three clusters: two homogeneous and one hybrid (dashed lines). Agents in the hybrid group form a bi-partite network and interact only with those of opposite type (for visual clarity links with less than 1\% interaction probability are not shown). \textbf{b)} Each bar represent one simulation containing 100 agents. Bars are sorted by the size of hybrid group. 87\% contain a hybrid group (S1R2-S2R1) and many simulation runs contain all three groups.
        }
    \label{fig:panel2}
\end{figure}

We find that distinct signaling groups emerge within a single population, and that dynamic networks alone are sufficient to produce this result (Fig.~\ref{fig:panel2}a). Agents converging on a signaling strategy maximize payoff by learning to visit only those with complementary strategies. This leads to the emergence of multiple groups of agents that use appropriately complementary signaling strategies. Each group uses either one of the homogeneous signaling systems or the hybrid solution with appropriate within-group network structure. Agents employing the complementary strategies thus form coherent clusters that interact predominantly among each other and rarely outside the group. Often, all three groups appear within a single population (i.e., a ``seed'' of the simulation; Fig.~\ref{fig:panel2}b). Distinct groups are able to form within the same population because subgroups of agents use network learning to update their visiting preferences and concentrate their interactions amongst each other. No agent in any of our 1,000 seeds adopted a primarily pooling strategy.

The most striking aspect of endogenous group formation is the frequent formation of hybrid (S1R2-S2R1) groups (Fig.~\ref{fig:panel2}b)): these groups appear in 87\% of seeds. This result is possible because network learning allows for directed preferential interaction and allows agents using the two complementary hybrid strategies to visit each other. Simulations show that the presence of hybrid groups becomes overwhelmingly likely as we generalize the game. In a $3\times3\times3$ game hybrid communication becomes more common than homogeneous (see {\it Supplemental Information}, Fig. S3). Inefficient pooling strategies are still a rare occurrence. On average, 71\% of agents are part of a hybrid group, 28\% a homogeneous group, and only 1\% employ a pooling strategy. The emergence of distinct signaling groups also reliably occurs if if one state of the world is more likely than the other. This is robust to changes in population size, discounting, and tremble rates (see {\it Supplemental Information}).

Finally, our analysis reveals important properties of hybrid groups. There is a strong bidirectional dependence between hybrid groups and the diversity of the other groups that emerge. The presence of a hybrid group increases the probability that at least one other signaling system survives in the population ($\chi = 35.4; p < 0.001$). The presence of both homogeneous group types increases the probability that a hybrid group forms ($\chi = 67.8; p < 0.001$). The two homogeneous group types rarely coexist without a hybrid group also being present (only 0.4\% of seeds). 

The network structure also differs between the two groups. Agents using one of the homogeneous signaling groups form regular networks, while agents belonging to a hybrid group form bipartite networks. In hybrid groups, link weights will only concentrate between agents using distinct complementary strategies forming a close resemblance to bipartite networks: a network in which only nodes of different types are connected. Agents in hybrid groups have some connections with agents who employ the same strategies, but these links are always extremely weak. 

To further explore the differences between hybrid and homogeneous signaling groups we analyze how information diffuses across the networks, how much outside observers (i.e., eavesdroppers) can glean from communication in the different groups, and how na\"ive agents behave when joining diverse populations. Information diffusion is equally effective, though hybrid groups diffuse information faster initially, but are slower to reach every member of the group. The bipartite networks have lower clustering which increases the short-term efficiency of information diffusion \cite{Vespignani2008DynProcesses}. The bipartite network structure of hybrid groups also plays a crucial role in how much information an outsider observer can glean. In a  homogeneous system the mutual information between signal and state or action is one, whereas the mutual information between signal and state or action in a hybrid group is zero. If an outsider observes a signal from a hybrid group, they cannot predict which state was observed or which action will be taken, making the group indecipherable to eavesdroppers who cannot also observe the state and the action. The behavior of na\"ive agents determines whether the more complex learning problem for hybrid groups---agents must learn both signaling strategy and bipartite network structure---makes it more difficult to join that group. In some circumstances na\"ive agents join hybrid groups as frequently as homogeneous groups: the probability of joining a hybrid vs.~homogeneous group is proportional to their share of the population, though this depends on network learning speed. These results, in addition to robustness analysis of parameter ranges, are detailed in the {\it Supplemental Information}.

\section*{Discussion}

Determining how order emerges from randomness is crucial to understanding complex systems. Past work on common interest signaling games, such as the Lewis sender-receiver game we use in our model, has shown that if near perfect communication is achieved, a population converges on a single signaling system. Our work shows the reliable emergence and stability of groups, including those using the hybrid solution to achieve near perfect communication in populations with endogenous network evolution. The dynamic network structure makes the hybrid solution viable by changing two features of standard approaches. First, agents can solve communication problems by finding compatible partners rather than experimenting with different signaling strategies. Second, it allows non-transitive interaction structures where agents can avoid, almost entirely, interacting with the partners of their partners resulting in bipartite network structures. These results present intriguing avenues for future work using empirical data and a deeper theoretical analysis. Another possibility involves generalizing our results for common interest signaling games to games where interests may conflict \cite{godfrey2013communication,huttegger2010dynamic}.

Our model predicts that we will only see hybrid signaling systems in networks that have non-transitive interaction structures which do not require strategies to be reflexively successful. Bipartite networks have these features. Our simulations show that hybrid signaling invariably co-evolves with bipartite network structures within the sub-graph. As these clusters form alongside unipartite clusters, overall network measures can be misleading and obscure the non-transitive interaction structure within hybrid clusters. This may make it difficult to recognize hybrid signaling in many social and natural systems. Analysis of sub-graphs can be used to identify bipartite clusters. These clusters could then be used as candidates for finding hybrid signaling groups within larger populations. Bipartite clusters are unlikely within standard signaling systems, despite this being a viable solution. Thus, our model predicts that when network evolution is driven by signaling success, bipartite structure is evidence of hybrid signaling.

There are ways to test if a group employs a hybrid signaling system. First, one can examine the diversity of signals and responses. Testing if diverse signals produce the same response, or if diverse responses can result from the same signal, would provide evidence for the presence of hybrid signaling. Second, experimental manipulation of the network can test whether two indirectly connected agents can successfully coordinate when a direct connection between them is created instead; if they fail, this may indicate a possible hybrid cluster.

Finally, our results contribute to research on how graphs evolve over time \cite{leskovec2005graphs,newman2003structure,foley2021avoiding}, providing an alternative approach that does not depend on growing graphs by modeling how new nodes attach to old nodes through processes like preferential attachment \cite{barabasi1999emergence}. In particular, our model can generate graphs with multiple communities, a pattern observed in many real world networks \cite{leskovec2005graphs}.

\printbibliography  

\cleardoublepage
\beginsupplement
\setcounter{section}{0}
\appendix
\setcounter{figure}{0}
\setcounter{table}{0}
\setcounter{equation}{0}
\setcounter{page}{1}
\renewcommand\theequation{\thesection .\arabic{equation}}
\rfoot{\thepage \hspace{1pt}/11}
\begin{refsection}

\section*{Supplementary Information}

\subsection*{Additional methods}

To support our analysis of the the coevolution of interaction structure and strategy in the Lewis Sender-Receiver signaling game \cite{lewis1969convention}, we investigate a more complex $3\times3\times3$ game, model the learning process of a na\"ive agent entering a population of established signaling groups, and examine a simple contagion process on the emergent network and strategy preferences formed in our first model to investigate information diffusion capacity. All model parameters are summarized in (Tab.~\ref{tab:Parameters}).

\paragraph{Extension to a more complex game.}
To investigate the robustness of our results, we also study a more complex $3\times3\times3$ game. Partner and strategy selection and updating occur exactly as described in the $2\times2\times2$ game. But there are now 3 states, and therefore 3 signals and 3 actions from which agents must select. This means instead of 16 sender-receiver strategy pairings, agents must use reinforcement learning to select from $(3^3 * 3^3)$ 729 possible strategies. The new $729x729$ payoff matrix is based on expected interaction payoffs under 3 equally likely states of the world. We initialized our model with the same baseline parameters as our 2 state game, and run each simulation for 1 million time steps.

\begin{table}[] \centering
\begin{tabular}{llr}
\toprule
\textbf{Symbol} & \textbf{Parameter Description}     & \textbf{Values} \\
\midrule
$N$               & Population Size                    & 20, 50, 100             \\
$w_{ij}^0$      & Initial Link Weights                         & $19/(N-1)$           \\
$s_{n}^0$      & Initial Strategy Weights                         & 5           \\
$\delta$        & Discount Factor                    & 0.01, 0.02           \\
$\epsilon$      & Error Rate                         & 0.01, 0.02           \\
$NLS$               & Network Learning Speed           & 0.1, 1, 10, 100              \\
$SLS$               & Strategy Learning Speed           & 1            \\
$T$               & Timesteps                          & $[0-1e^6]$\\
\bottomrule
\end{tabular}
\caption{Parameters and baseline values.}
\label{tab:Parameters}
\end{table}

\begin{mycomment}
\subsection*{Analytic Model (in progress...)}

Here we introduce a simplified analytic model with discrete network ties and pure-strategies that corroborates our simulation results and demonstrates the stability of each type of signaling system identified.

Suppose we have $N$ agents, each agent has a single outgoing link to another agent and can employ a single pure strategy for each round. Suppose that each round agents update both their link and strategy by selecting a \emph{best-response}, choosing randomly among best-responses if there are more than one. We will suppose that each round, first links are formed, then strategies updated. Strategies and links are initially assigned at random.

Agents using S1R1 will link to other S1R1 agents (if present). Likewise with agents using S2R2. Agents using S1R2 will link to agents using S2R1 (if present), and vice versa. 

Provided there are at least two members of any group, the group will remain stable.

Alternative strategies (not S1R1, S2R2, S1R2 and S2R1) will not be stable, as there are always multiple strategies that do equally well (and strategies are chosen at random among best responses). One of the four signaling strategies is always among the best responses (for any non-signaling strategy). Therefore, all non-signaling agents will eventually adopt a signaling strategy. Once they do, they will link to another agent with a complementary signaling strategy.

If all agents are employing a signaling strategy and are linked to one complementary agent, they will not alter their strategies.

\end{mycomment}

\paragraph{Na\"ive agent simulations}

To test how a na\"ive agent entering a population of established signaling groups would behave, we explored different initializations of our first model. To ensure robust results and have the ability to precisely control the scenarios, we use stylized starting positions rather than using ``real'' outcomes from simulations. The initial network and strategy weights were chosen to approximate a population with multiple stable signaling groups. For example, any agent in a S1R1 group was initialized with a weight of 10 on the S1R1 strategy and a weight of 0.1 on all other strategies. Each agent also selected a preferred partner with a complem strategy at random and assigned a network weight of $90$ to that agent. All other network connections received a weight of $0.1$. In the case of hybrid groups, half of agents placed the majority of their strategy weight on the S1R2 strategy and the other half on the S2R1 strategy. Each then choose at random a partner with a complem strategy and assigned a network weight of $90$ to that agent, and a weight of $0.1$ for all other connections. These initialization weights roughly approximate the patterns formed in our original simulations. Using this approach we can systematically vary over different configurations of populations of 100 agents. With this starting setup, we add a single additional na\"ive agent to the population. The na\"ive agent is initialized with uniform network and strategy weights identical to the original simulation. We then simulate the signaling game for 50,000 time steps.


\paragraph{Diffusion model simulations}
To test information diffusion capacity of the emergent signaling groups, we created a simple contagion model. Intuitively, we take the final state of the simulation of the main game ($t=1e^6$) as our starting point, and simulate an information diffusion process \textit{on this final state (network and strategy)} without further updating network or strategy weights. Based on the network and strategy weights learned up to that point, the population communicates to diffuse information through the population. That is, we simulate a kind of susceptible-infected diffusion process where interaction depends on a weighted network and transmission depends on successfully completing one instance of the Lewis Sender-Receiver game between the two agents with their respective strategy weights (instead of the typical probabilistic transmission used in $SI$ models). We then analyze the diffusion process with regard to the fraction of the population that has ``heard of the information'' after a certain number of time steps. We simulate the diffusion process separately for each individual group formed across all seeds of our original set of simulations. Groups are defined using a network thresholding approach. In this model, a single agent in each group is seeded with information that needs to be passed on to other agents. Information is only passed from one agent to another if the two agents communicate ``successfully'' according to the payoffs of the Lewis Sender-Receiver game. That is, information is only passed between agents when they correctly coordinate their signals and actions ($\pi_{ij}=2$). If two agents communicate successfully, we consider the information as having passed on. The order of interactions in each round is randomized, and any newly ``infected'' agent can spread the information further within a round. Once an agent has received the information they remain ``infected'' and continue to spread it to other agents in the network (i.e., they do not recover and stop spreading). For the purpose of studying within-group diffusion speed, we restrict interactions to those within the group. If an agent selected an interaction partner outside of their own group, no interaction took place. Simulation ended when every agent in the group had obtained the information. We show results as the fraction of agents who had received the information after each time step. We report results averaged across 50 trials. Each trial randomly selects one agent as the seed of the of the information.

\subsection*{Related work}
We review related work in three key areas: (1) the Lewis Sender-Receiver game; (2) other similar language games; and (3) statistical physics of social dynamics.

First, previous work on the Lewis Sender-Receiver game has demonstrated that dynamical models are necessary to explain how a single signaling system can be selected amongst equivalent alternatives \cite{skyrms2010signals}. Using this insight, Lewis Sender-Receiver games have been used to study the evolution of conventional meaning in a wide variety of contexts. These range from evolutionary models based on biological reproduction \cite{hofbauer2008feasibility,skyrms2014evolution}
, to models of cultural evolution via imitation \cite{zollman2005talking, franke2020vagueness}, to studies of pairs or groups of individual learners \cite{barrett2009role, argiento2009learning, skyrms2010signals, hu2011reinforcement}. Studies have examined the robustness of the evolution of meaning by varying assumptions like population size \cite{pawlowitsch2007finite}, randomness of pairing \cite{skyrms2014evolution}, initial strategy randomness \cite{LaCroix2018}, and the probability of states \cite{hofbauer2008feasibility}. For the most part, these studies find a strong convergence to successful signaling across a wide variety of settings, learning rules, and evolutionary dynamics \cite{huttegger2014some,LaCroix2020}. There are some complications, however. Successful communication is not guaranteed to arise in settings involving more than two states, signals and acts \cite{Barrett2006, huttegger2010evolutionary}, nor is it guaranteed if the states are not equiprobable \cite{hofbauer2008feasibility}. 

The existing work most closely related to our own are models that incorporate some spatial or networked pattern of interaction. Such work includes learning to signal with neighbors on a grid \cite{Zollman2005}, and neighborhood interactions in more complex network structures \cite{Wagner2009,muhlenbernd2012signaling}. There are also spatial models where agents interact probabilistically based on distance \cite{muhlenbernd2011learning}. This demonstrates the robustness of successful signaling in systems that are not fully local nor fully random. While the coevolution of strategy and network has gained prominence in the class of social dilemma games \cite{FoleyEtAl2018,foley2021avoiding,Fulker2021}, no existing work has examined its relevance in the Lewis Sender-Receiver game \cite{perc2010coevolutionary}. Moreover, to our knowledge, no study on the Lewis Sender-Receiver has explicitly discussed the evolutionary viability of hybrid signaling strategies.

Second, in addition to the Lewis Sender-Receiver game, a variety of other games have been used to study the evolution of language and signals. One of the earliest was the Evolutionary Language game, where successive generations of agents update a speaking and listening matrix based on a sampling of others in the population \cite{Nowak1999Bio, nowak1999evolution}. Another common model of language evolution is the naming game \cite{Baronchelli2006, Baronchelli2008}. In the simplest version of the game agents keep an inventory of words, a speaker randomly selects one word from their inventory, and if the word is also in the hearers inventory, both agents delete all other words. The goal of the game is for all agents to have a single shared word in their inventory. The Category (or Colour) game is a slightly more complex version of this game but is informed by the universal need for cultures to create a naming system for colors \cite{Steels2005, Baronchelli2010}. In this game, agents must learn to subdivide a continuous spectrum into discrete nameable units, in addition to reaching a consensus on the names themselves. In each of these games, however, one agent must send information about a stimuli to another listening agent. And just as in the Lewis Sender-Receiver game, agents must gradually learn a successful communication strategy from interactions with others. To our knowledge, no one has studied the coevolution of strategy and network in these related language games. 

Third, our research relates to work on the statistical physics of dynamical systems. This approach is often grounded in a single question: how do the interactions between agents create order out of an initial disordered situation \cite{Castellano2009}? Research in this field has also been interested in signaling and language dynamics. Models of language dynamics can be divided into two categories: sociocultural and sociobiological approaches. Sociocultrual approaches treat language as a continuously evolving complex dynamic system. In this setting improvements in strategy improve an agents communication ability \cite{Ke2008}. Our model can be seen as an instance of this, where agents gradually learn new strategies which increase signal coordination and improve payoffs. The community of agents collectively develops and refines an effective communication system. Sociobiological approaches to language dynamics treat successful communication as a selective advantage. In this setting improvements in strategy improve an agent's reproductive fitness \cite{nowak1999evolution}. This means successful communication strategies lead to more offspring using the same communication strategies. Several models have tested the hypothesis that communication strategies are innate and passed genetically to offspring \cite{Huford1989}. This work aligns with the nativist approach to language acquisition \cite{Chomsky1965}. One such seminal paper explored the evolution of Saussurean signs \cite{Huford1989}. A Saussurean sign describes a bidirectional link between a concept and a sound (or other form of communication). Our work relates to the distinction between Saussurean and non-Saussurean communication. The homogeneous signaling we describe is an instance of Saussurean communication, while our hybrid signaling is an instance of non-Saussurean communication.

\subsection*{Group formation and network structure}\label{SI:EndogenousGroupFormation}\label{SI:Thresholding}
\begin{figure}[t!] \centering
    \includegraphics[width=.8\linewidth]{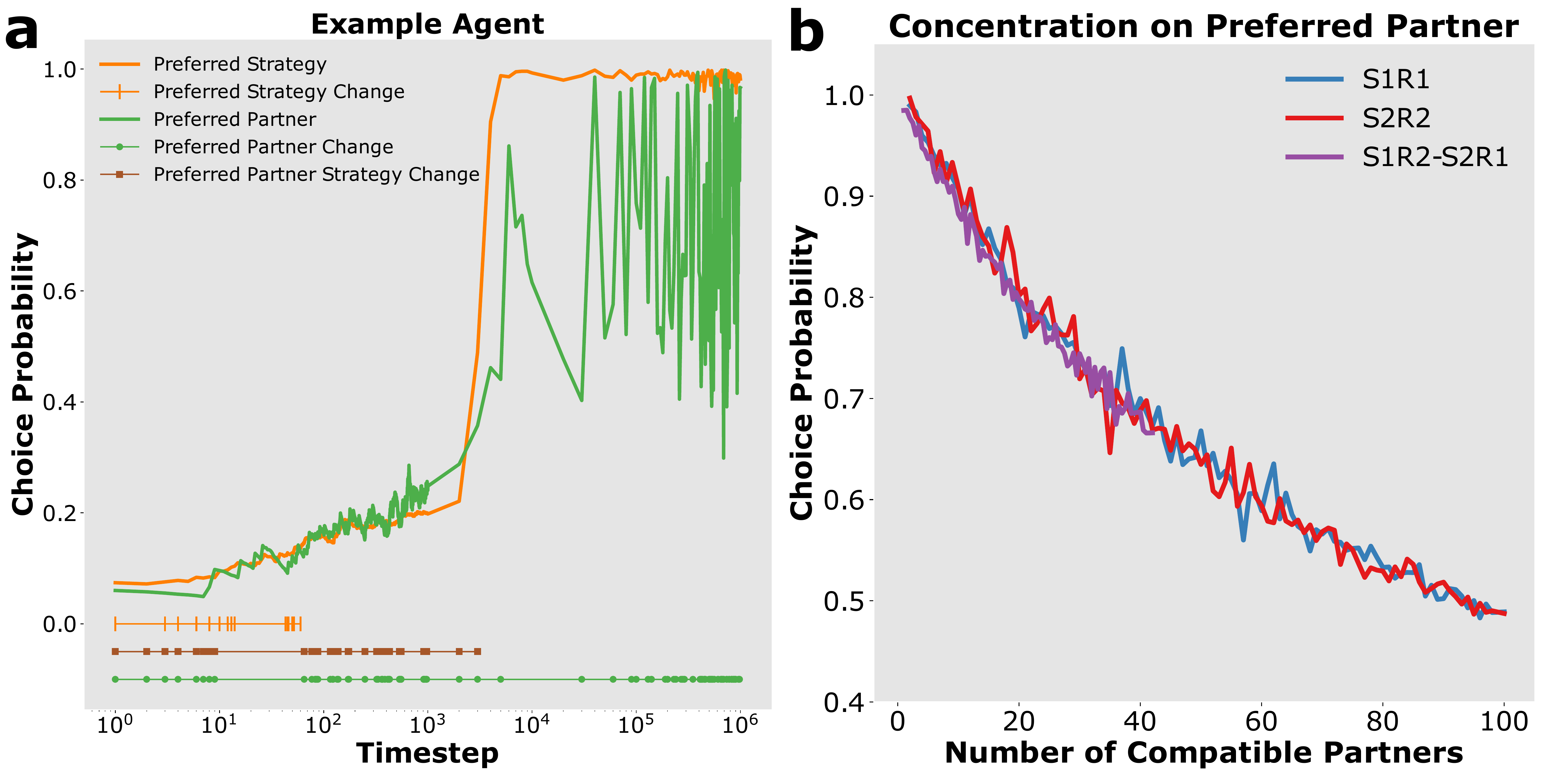}
    \caption{
        \textbf{Network Properties. a)} Despite most agents having a strongly preferred partner at any point in time, the identity of this partner frequently cycles among compatible partners. In most cases agents settle on a preferred strategy before identifying compatible partners for that strategy. \textbf{b)} While the structure of hybrid and homogeneous networks differ, many of the their properties are the same. Agents in each type of group maintain similarly large average interaction probabilities with their preferred partner.
        }
    \label{fig:panel4}
\end{figure}

Over time agents identify the strategies that provide maximum earnings (Fig.~\ref{fig:panel4}a). Simultaneously agents strengthen network ties with favorable partners. Eventually agents reach a state where a large share of their link weight rests on a single preferred partner. Yet this network configuration is not static. Once agents settle their strategies, they shift their preferred partner in a random walk process amongst complementary partners (Fig.~\ref{fig:panel4}b) \cite{SkyrmsPemantle2000}. The networks continue to exhibit temporality \cite{li2017fundamental}. On average agents placed 98.7\% of their strategy weight on their preferred strategy. On average 99.7\% of an agent's network weight is allocated to others who primarily use a complementary strategy. 

This feature of agent-agent connections permits the emergent property of group formation to be recovered from simple network thresholding. To recover the groups formed in each seed of our simulation we used a normalized threshold of $0.1$---links representing an interaction probability of less than 10\% were dropped and the groups are the subsets of agents who remain (weakly) connected via the network after this process. Because the agents in our model concentrated almost all of their outgoing weight on complementary strategies (99.7\%), the network groups recovered were highly robust to changes in thresholding value and subsetting methods. In most cases an agent will first settle on a final preferred strategy, then adjust their network weights accordingly. This order occurs in about two-thirds of cases, while in one-third of cases an agent first settles on a preferred partner and adjusts their own strategy to be complementary.

The groups recovered via our thresholding approach were almost always composed of agents using complementary strategies. Occasionally, separate groups using the same signaling strategy form in a single seed. This occurred for at least one group type in 5.8\% seeds. This result is explained by the random walk process of shifting link weights within an established signaling group. As link weights within a signaling group shift, a subset of these agents may become isolated. But this is a random artifact of the moment we decided to stop our simulation and threshold the network. Because agents using complementary strategies rarely form separate network groups, the choice to define a group based on strategy or network weights does not significantly affect our results.

We find that no agent in any of our simulations adopted a primarily pooling strategy. Thus, signaling is nearly perfect within each group. Without network learning, groups do not form; with faster network learning, groups form more easily and the percentage of populations in which all three signaling systems emerge increases.

\begin{figure}[t!] \centering
    \includegraphics[width=.8\linewidth]{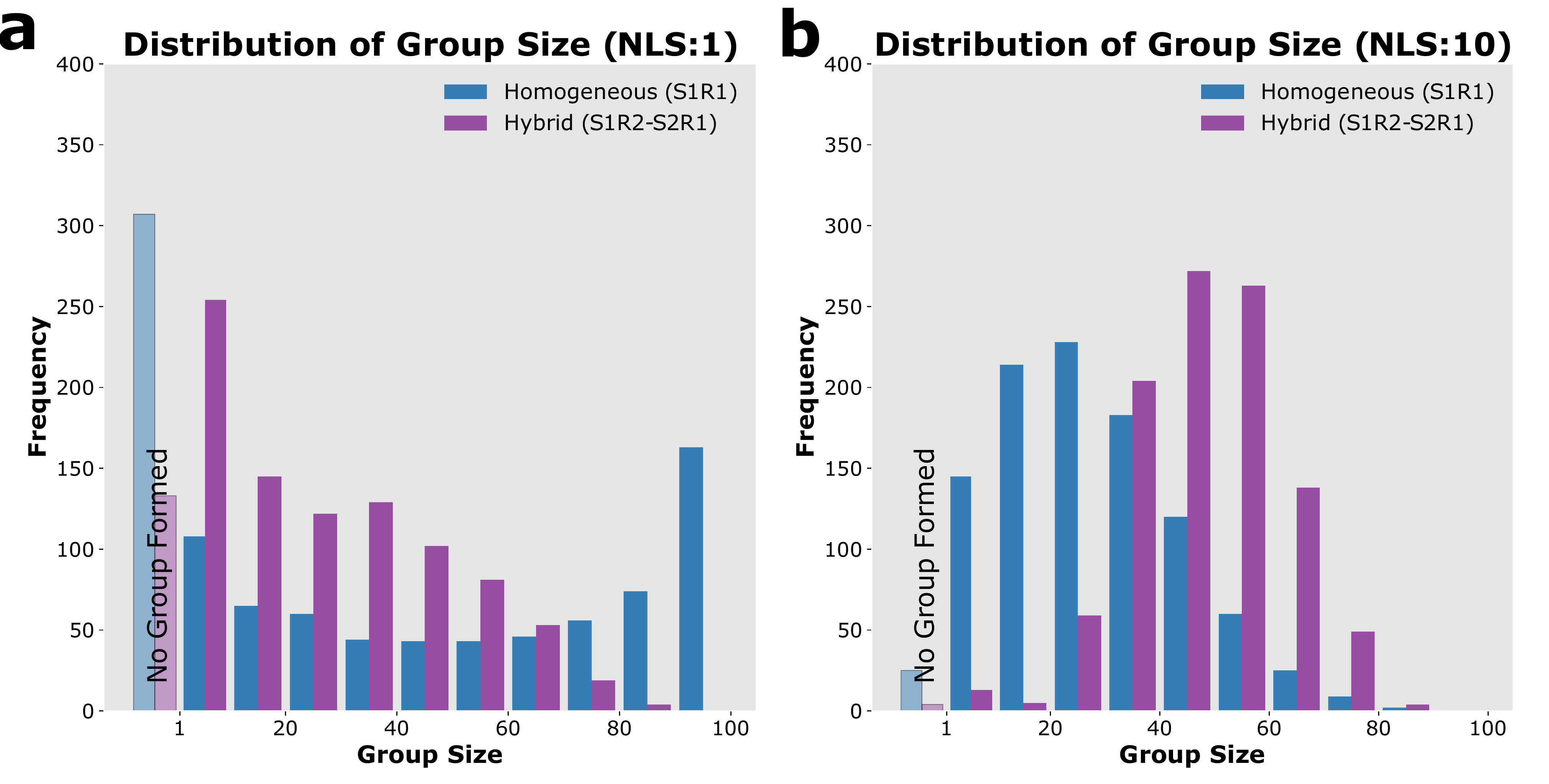}
    \caption{
        \textbf{Group Size by Type. a)} The distribution of group sizes demonstrates that while the presence of hybrid groups is more robust, they tend to be smaller in size than their homogeneous counterparts. Furthermore, no population converges to a single hybrid group, despite this being a relatively common outcome for homogeneous groups. \textbf{b)} Faster network learning increases the size of hybrid groups who must overcome additional constraints in their network structure. Faster network learning also decreases the likelihood that any group type fails to form in a given population. This is because agents are able to more quickly sort themselves into insular groups enabling more strategy diversity. For visual clarity we plot only the distribution of S1R1 groups and not S2R2 groups, because as expected their distributions are nearly identical. 
        }
    \label{fig:panel5}
\end{figure}

With respect to the composition and sizes of emergent groups, network learning speed (NLS) has a large effect. When NLS is large, agents are able to adjust their preferred partners more quickly. When network learning speed increases, the average size of hybrid groups also increases (Fig.~\ref{fig:panel5}a,b). The hybrid group benefits from faster NLS because it faces a more complicated network learning problem. Agents using a hybrid strategy have about half as many potential complementary partners to that of a similarly sized homogeneous group. In general, faster network learning allows subsets of the population to insulate themselves more quickly. This means all three types of signaling groups appear within a population with greater frequency ($94.3\%$ vs.~$40\%$ when network learning speed is $10$ vs.~$1$). With faster NLS, no population converges on a single group type.  

\subsection*{Prevalence of hybrid groups as state space increases}


Hybrid groups become more common in more complex games because the number of pairwise combinations for complementary hybrid solutions increases more quickly than homogeneous solutions. Therefore the share of agents in a hybrid group increases as the complexity of the game increases further (e.g., a $4\times4\times4$ game). Given sufficiently fast network learning speed (NLS) agents are roughly equally likely to stumble upon any of the perfectly coordinating signaling strategies. In the two-state game there are two homogeneous strategies and one pair of complementary hybrid strategies. As expected, more agents adopt a homogeneous strategy compared to a hybrid strategy. In the three-state game there are six homogeneous strategies and 15 pairs of complementary hybrid strategies, leading the majority of agents to adopt a hybrid strategy. In fact as the number of $N$ states grows, the share of perfectly coordinating strategy pairings that are homogeneous approaches $0$. The number of homogeneous strategies in an $N \times N \times N$  game is $N!$. This is because there are $N!$ ways to assign a unique signal to each state (i.e., a separating strategy). The number of hybrid pairings in an $N$ state game is $(N!(N!-1))/2$. Hybrid strategies are composed of two of the $N!$ unique separating strategies, one for sender and one for receiver strategy. The complement of any hybrid strategy is the same two unique separating strategies, but a flipped assignment to sender and receiver strategy. In this way when agent one's sender strategy is matched with agent two's receiver strategy the exchange is akin to a homogeneous interaction. The same is true when agent two's sender strategy is matched with agent one's receiver strategy, but this time a different homogeneous pairing is used. Therefore, the ratio of homogeneous group types to hybrid group types in an $N \times N \times N$ game is given by $\frac{N!}{(N!(N!-1))/2}$ which approaches 0 as $N\to\infty$.

\begin{figure}[t!] \centering
    \includegraphics[width=.4\linewidth]{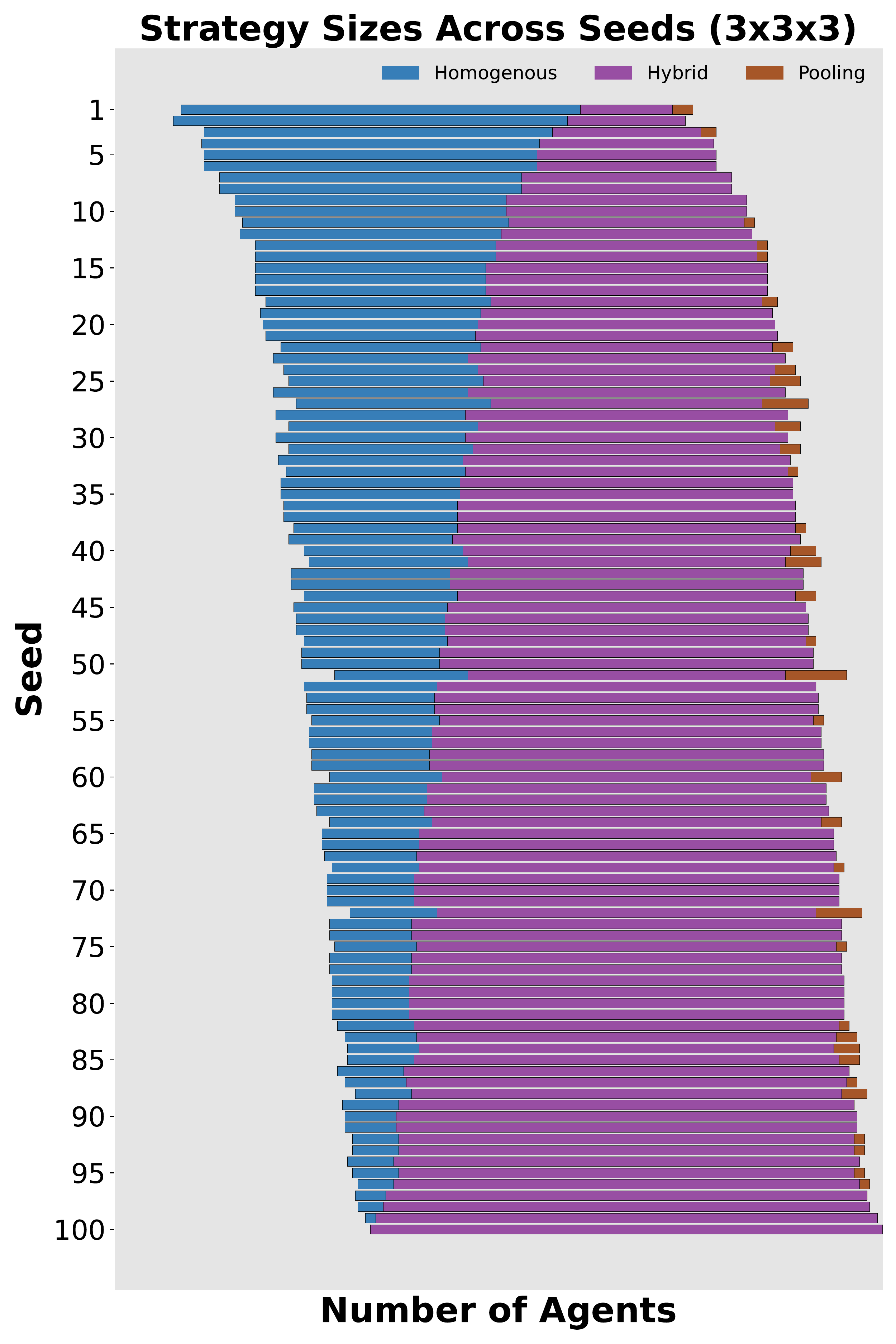}
    \caption{
        \textbf{Increased State Space} In the 3 × 3 × 3 game hybrid strategies become the most likely outcome and
        appear in every seed. Pooling strategies remain rare despite being a more common outcome for 3 state games on static
        networks. 
        }
    \label{fig:panel6}
\end{figure}

\subsection*{Hybrid groups promote diversity}

Hybrid groups have a positive relationship with the number of signaling group types formed within a seed because each hybrid strategy can partially coordinate with the two homogeneous strategies. Specifically, the two different homogeneous strategies receive a payoff of $0$ when interacting, while hybrid and homogeneous strategies earn a partial payoff of $1$ when interacting with each other. The effect of this partial payoff is that the hybrid strategy is still reinforced even in an environment where many interactions occur with different (homogeneous) strategies. The result of this diversity advantage is that hybrid groups form more consistently (87\% of seeds vs.~69\% and 70\% for homogeneous types). Similarly, this partial payoff also reinforces either homogeneous strategy in an environment where many interactions occur with hybrid strategies. This means smaller emerging groups are less likely to be absorbed into a larger hybrid group in comparison to a larger homogeneous group.

In order to verify a bidirectional causal relationship between hybrid groups and diversity we ran a set of controlled experiments. In the first, we seeded each simulation with an established homogeneous group of four agents (Fig.\ref{fig:panel9}a). And in the second, we seeded each simulation with an established hybrid group of four agents (Fig.\ref{fig:panel9}b). All other model parameters remained the same. We find that a second signaling group emerges significantly more often in the second set of simulations seeded with a hybrid group (46 vs.~99; $\chi = 35.4; p < 0.001$).This demonstrates that the presence of a hybrid group, rather than a homogeneous group, increases the likelihood of other groups forming. Finally, we seeded a set of simulations with a dyad of two agents from each of the two homogeneous types (Fig.\ref{fig:panel9}c). In this case, we find that a hybrid group forms significantly more often in the simulations seeded with both homogeneous types compared to the first set seeded with only a single homogeneous type (86 vs.~45; $\chi = 67.8; p < 0.001$). This demonstrates that hybrid groups are more likely to form in populations with greater group diversity. 

Another consequence of the lack of a partial payoff between the two homogeneous strategies is that these systems rarely appear together in the absence of a hybrid group (0.4\% seeds). This is due to a positive network externality. This describes a situation where the value each individual receives increases with the number of others who have already adopted the same behavior. In our model, the larger of any two nascent homogeneous groups will have an expected earnings advantage (e.g., a positive network externality). This advantage is due to randomness and tremble in the network learning process being more likely to yield a complementary partner strategy for agents in the larger group. The additional reinforcement provided through these earnings will push most agents to adopt the strategy of the larger group. Even after groups are solidified, larger groups benefit from higher average earnings (Fig.~\ref{fig:panel9}d). The coexistence of S1R1 and S2R2 groups becomes significantly more likely when a hybrid group is also present (39.0\% of seeds contain all three). When all three types of groups coexist, hybrid groups form a buffer between the distinct homogeneous strategies, and reduce the frequency of their cross-interactions. This helps protect the smaller of the two homogeneous systems from being invaded by the other. The existence of the hybrid group creates a boundary in the network that functions analogously to a geographic boundary in biological speciation.

\subsection*{Properties of homogeneous and hybrid groups}

\paragraph{Network structure.}
Another key emergent difference between the hybrid and homogeneous groups is their network structure. Across both homogeneous and hybrid groups we find that agents have a skewed network weight distribution with most of the weight concentrated on a single interaction partner. However, who this preferred interaction partner is, is not stable but changes from time to time. Within homogeneous groups, these strongly weighted network links are distributed randomly throughout the group because agents can perfectly coordinate with any group member equally well. In hybrid groups, link weights will only concentrate between agents using distinct complementary strategies forming a close resemblance to bipartite networks: a network in which only nodes of different types are connected. Agents in hybrid groups have some connections with agents who employ the same strategies, but these links are always extremely weak. This difference in network structure also means hybrid and homogeneous groups are effected differently by changes in network learning. Namely, the share of hybrid agents in the population is dependent on the network learning speed.

\begin{figure}[!hb] \centering
    \includegraphics[width=.6\linewidth]{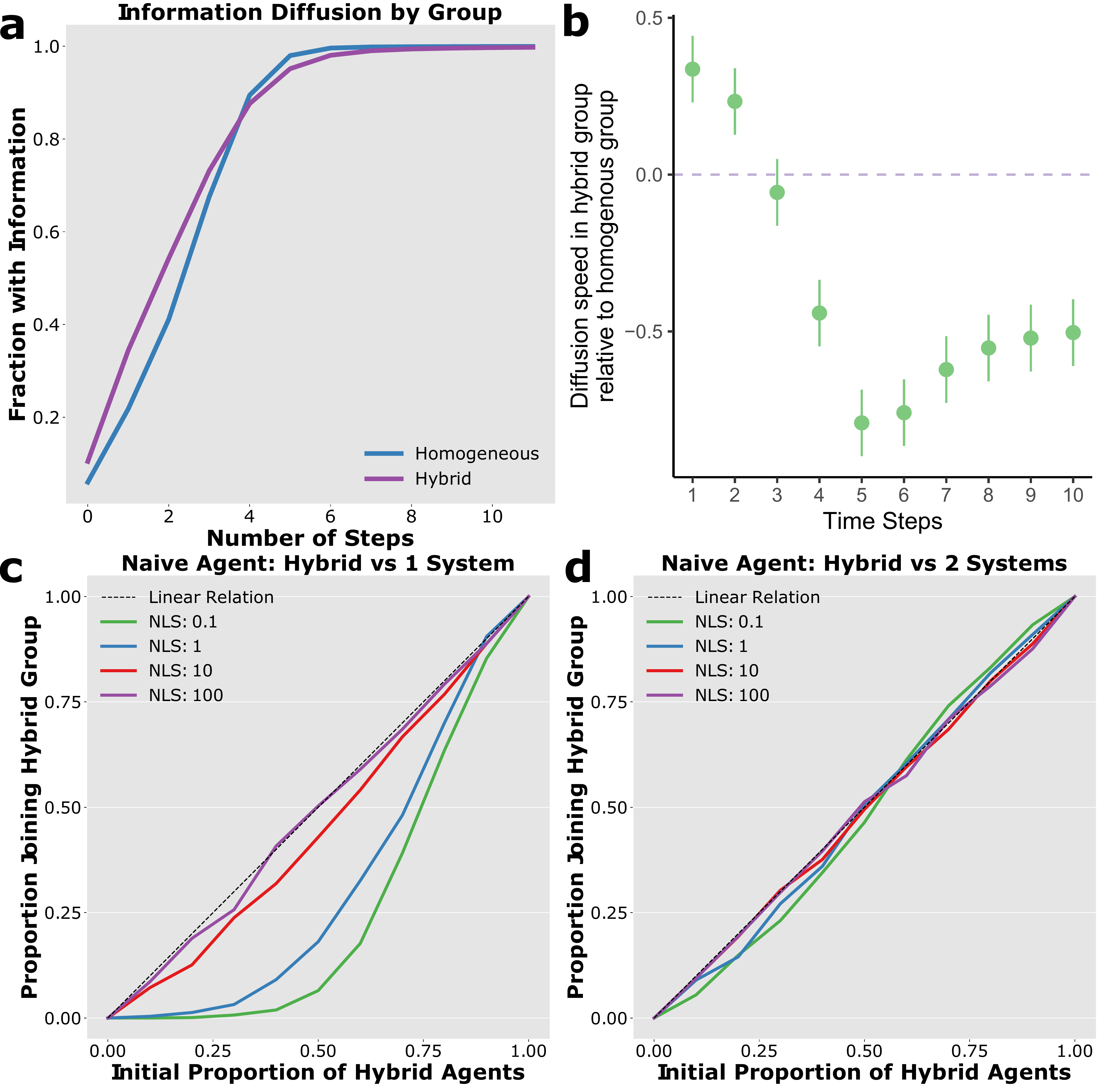}
    \caption{
        \textbf{Signaling Group Properties. a)}  All three types of signaling groups diffuse information in a similar speed and pattern. Hybrid groups, however, initially spread information faster than their homogeneous counterparts before becoming slower at later time steps. \textbf{b)} A regression analysis controlling for group size confirms the robustness of this result. \textbf{c)} Naive agents entering an established population with one hybrid group and one homogeneous group join the hybrid group at a lower rate than its initial share of the population. As network learning speed is increased naive agents can more easily find a network solution compatible with using a hybrid strategy and this relationship becomes linear. \textbf{d)}  When the initial set of homogeneous agents is instead split equally between the two homogeneous types, naive agents join the hybrid group a rate proportional to their initial share of the population. NLS no longer influences which group naive agents join because all signaling strategies must find an equally difficult network learning solution.
        }
    \label{fig:panel7}
\end{figure}

\paragraph{Information diffusion.}
A key function of any signaling group is to quickly and efficiently transmit information. We find that both hybrid and homogeneous groups can effectively diffuse information, but differ in their diffusion speed over time (Fig.~\ref{fig:panel7}a) (see Methods). A regression analysis of the diffusion rate controlling for size reveals the differences between the signaling groups. Hybrid groups diffuse information faster initially, but are slower to reach every member of the group (Fig.~\ref{fig:panel7}b). The initial advantage of hybrid groups is due to their near bipartite structure reducing the frequency of redundant information sharing. For a hybrid agent, the friends (i.e. interaction partners) of my friend are not my friends. In other words, the bipartite network has low clustering which increases the short-term efficiency of information diffusion \cite{Vespignani2008DynProcesses}. This is not necessarily true for the networks formed by homogeneous groups. Our finding that overall diffusion is marginally slower in hybrid groups is in line with prior work showing that communication efficiency \cite{PhysRevLett.87.198701} in bipartite networks with balanced node classes decreases as the network becomes more strictly bipartite \cite{SinghEtAl2019}.

We use ordinary least squares regression to investigate differences in diffusion speed. Specifically, we investigate the rate of change (logit) of the proportion of the group that has been infected at each time step. Using the homogeneous S1R1 group as reference group, we contrast how the other homogeneous group (S2R2) and the hybrid group differ. As expected, the S2R2 group does not differ at all with all coefficients around $0.0$ and not statistically significant (coefficients omitted from table).

\begin{table}[h!]
\footnotesize
\begin{center}
\begin{tabular}{l D{.}{.}{5.5}}
\toprule
 & \multicolumn{1}{c}{Model 1} \\
\midrule
(Intercept)            & -1.93^{***} \\
                       & (0.03)      \\
 Hybrid          & 0.05        \\
                       & (0.04)      \\
 Homogeneous (S2R2)               & 0.02        \\
                       & (0.04)      \\
Signaling Group Size             & -0.01^{***} \\
                       & (0.00)      \\
 Hybrid $\times$ Step 1  & 0.34^{***}  \\
                       & (0.05)      \\
 Hybrid $\times$ Step 2  & 0.23^{***}  \\
                       & (0.05)      \\
 Hybrid $\times$ Step 3  & -0.06       \\
                       & (0.05)      \\
 Hybrid $\times$ Step 4  & -0.44^{***} \\
                       & (0.05)      \\
 Hybrid $\times$ Step 5  & -0.79^{***} \\
                       & (0.05)      \\
 Hybrid $\times$ Step 6  & -0.76^{***} \\
                       & (0.05)      \\
 Hybrid $\times$ Step 7  & -0.62^{***} \\
                       & (0.05)      \\
 Hybrid $\times$ Step 8  & -0.55^{***} \\
                       & (0.05)      \\
 Hybrid $\times$ Step 9  & -0.52^{***} \\
                       & (0.05)      \\
 Hybrid $\times$ Step 10 & -0.50^{***} \\
                       & (0.05)      \\
Time Step Dummies & \multicolumn{1}{c}{\textit{Included}}\\
Time Step Dummies $\times$ Homogeneous (S2R2) & \multicolumn{1}{c}{\textit{Included}}\\
\midrule
R$^2$                  & 0.89        \\
Num. obs.              & 25,476       \\
\bottomrule
\multicolumn{2}{l}{\scriptsize{$^{***}p<0.001$; $^{**}p<0.01$; $^{*}p<0.05$}}
\end{tabular}
\caption{Diffusion speed regression (coefficients for Fig.~5b).}
\label{table:coefficients}
\end{center}
\end{table}

\paragraph{Na\"ive agents.} Which signaling group would a newcomer join? Conceptually, it seems that a na\"ive agent would find it difficult to join a hybrid signaling group as that agent would have to learn both the signaling strategy and the bipartite network structure. We investigate this question by adding one additional agent with uniform network and strategy preferences to established signaling populations. We find that in some circumstances na\"ive agents join hybrid groups as frequently as homogeneous groups (proportional to their share of the population). In other cases, agents are more likely to join homogeneous groups. 

Specifically, there are two scenarios in which na\"ive agents join hybrid groups at a linear rate proportional to their initial share of the population. The first is when network learning speed (NLS) is sufficiently large (Fig.~\ref{fig:panel7}c). Faster network learning compensates for the relative reduction in the number of complementary partners available in hybrid groups due to their near bipartite structure. The second is when the non-hybrid share of population is split into two equally sized homogeneous groups using distinct strategies (Fig.~\ref{fig:panel7}d). NLS has no effect because homogeneous and hybrid strategies have the same number of complementary partners relative to their share of the population.

\paragraph{Mutual information.} What amount of information can an outsider observe about the communication of groups? We consider the amount of information observing a signal in a given group provides about the action that will be taken by the agents in that group. We find the bipartite network structure of hybrid groups plays a crucial role. In a perfectly efficient homogeneous system the mutual information will be one. This is because there is a perfect relationship between the signal sent and the action taken. However, the mutual information of a perfectly efficient and balanced hybrid group would be zero. Because no agreement is reached on the meaning of each signal, each time signal $s_1$ is sent half of the agents will take action $a_1$, and half will take action $a_2$. Therefore, if an outsider observes a signal from a hybrid group, it provides no information about which action will be taken. The low mutual information of hybrid groups has the benefit of making the group indecipherable to eavesdroppers who cannot observe the source of the signals and actions taken in the group.

\begin{figure}[!hb] \centering
    \includegraphics[width=.6\linewidth]{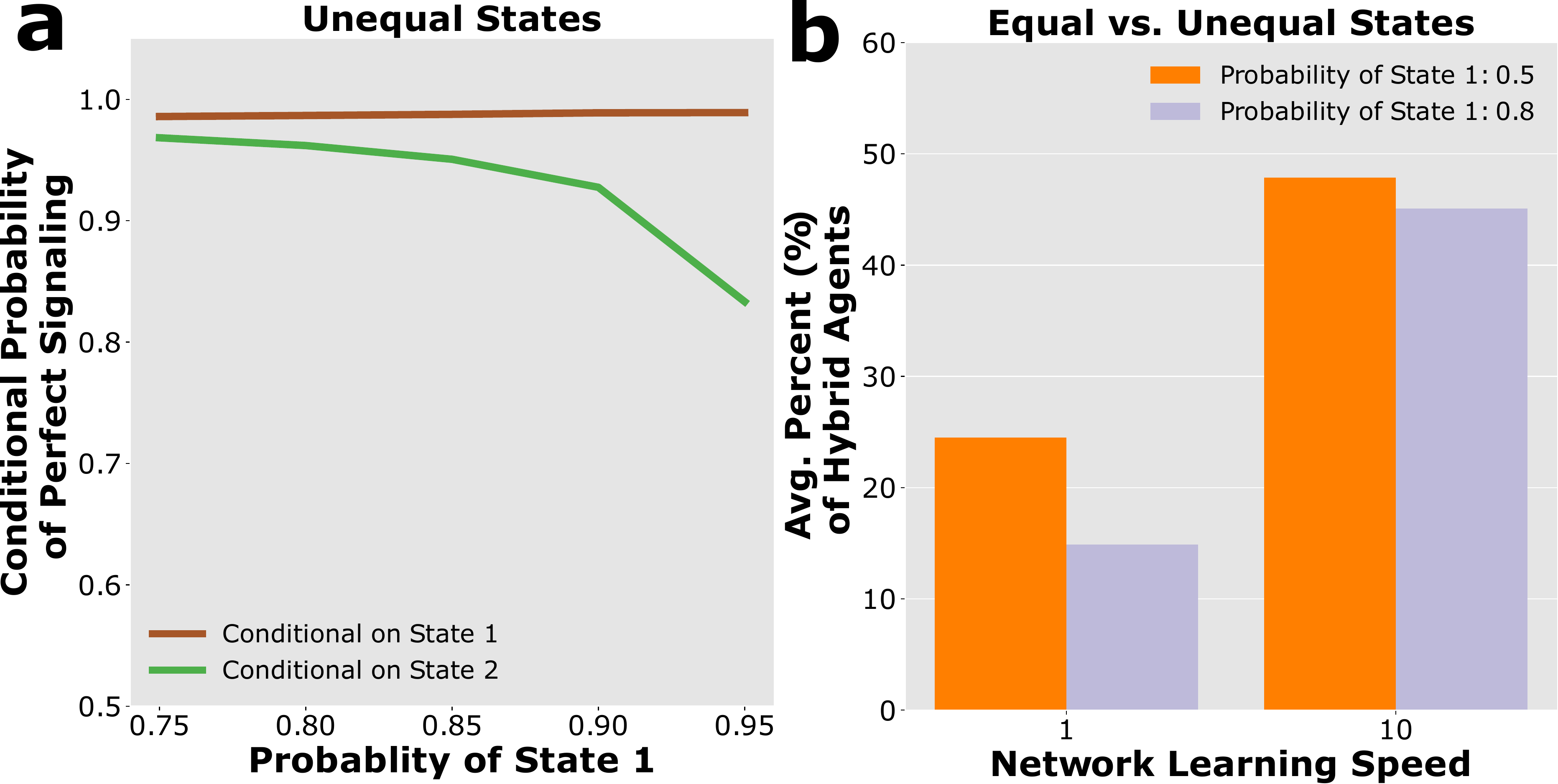}
    \caption{
        \textbf{Unequal States. a)} As the likelihood of state 1 increases (and the likelihood of state 2 decreases), agents still coordinate their signals to most often earn the maximum payoff in the unlikely state. \textbf{b)} When states are unequal the average percentage of agents who primarily use a hybrid strategy decreases. However, the hybrid strategy is still prevalent and increases in network learning speed can offset the effect of unequal states. 
        }
    \label{fig:panel8}
\end{figure}

\subsection*{Robustness to parameter changes}\label{SI:Robustness}
As is common in theoretical and (agent-based) modeling work, our study relies on certain assumptions and parameter choices. In this section, we explore the robustness of our findings to the most important parameter choices and assumptions we made. 

\paragraph*{Endogenous group formation when initial states are unequal.} How robust is the emergence of groups? As a robustness check we consider how our results of the $2\times2\times2$ game change when one state of the world becomes (much) more likely than the other. The strategy and network updating process of our model remains unchanged, only the expected frequency of states changes. Consider an agent who always takes action $a_1$ regardless of the signal they receive. If the likelihood of state $s_1$ is increased to $0.9$, the expected payoff this agent earns as a receiver increases from $0.5$ to $0.9$. 
Our examination of unequal states demonstrates that the formation of signaling groups using separating strategies is robust (Fig.~\ref{fig:panel8}a). Even in the case of extremely unequal states, agents still learn an effective signaling strategy for the unlikely state. This means agents most often avoid using a pooling strategy, even though these strategies can earn nearly the maximum payoff. In fact, only 0.19\% of agents adopt a primarily pooling strategy when $P(s_1) = 0.9$. We also find that hybrid groups, while less prevalent under unequal states, still consistently emerge, especially when network learning speed is high (Fig.~\ref{fig:panel8}a).

\begin{figure}[t!] \centering
    \includegraphics[width=0.8\linewidth]{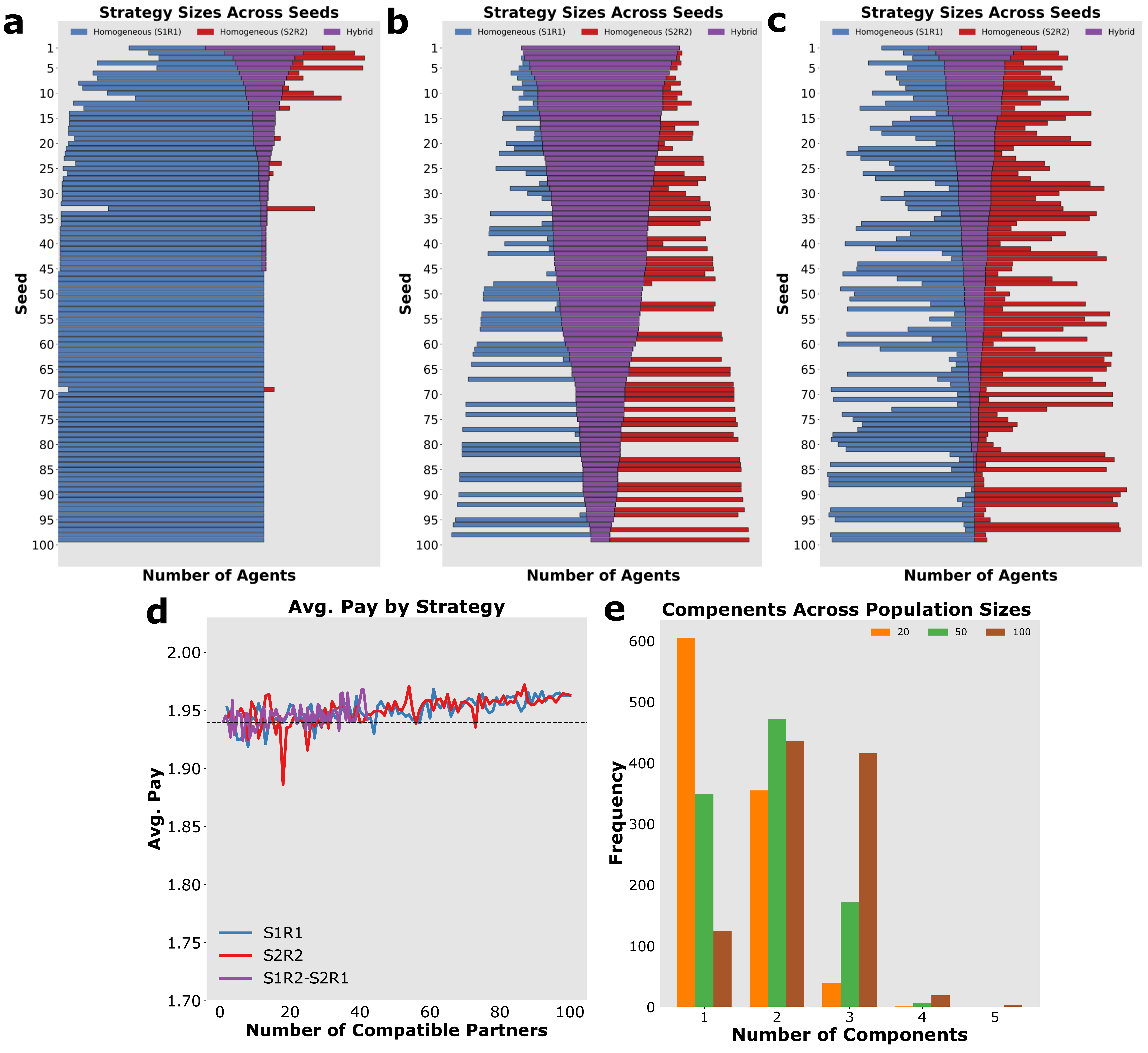}
    \caption{
        \textbf{Seeded Simulations, Network Effects, and Robustness. a)} When we seeded simulations with an established group of four homogeneous (S1R1) agents, a second type of signaling group emerged in less than half of seeds (46/100). \textbf{b)} When we seeded simulations with a group of four hybrid agents, a second signaling group almost always emerged (99/100). \textbf{c)} In simulations seeded with two distinct homogeneous groups of size two, hybrid groups emerge more frequently than in simulations seeded with only a single homogeneous type of size four (86 vs. 45). Together these three results support that hybrid groups both allow for, and are resilient to diversity. \textbf{d)} In the final timesteps of our model, agents belonging to larger signaling groups earn larger average payoffs. \textbf{e)} The formation of multiple network components (i.e., signaling groups) is robust even in small populations.
        }
    \label{fig:panel9}
\end{figure}

\paragraph{Population size.} Population size has a critical effect on evolutionary models as there are many known small population effects. We find that even small populations can support signaling group diversity (Fig.~\ref{fig:panel9}e). 

\paragraph{Discount and tremble rate.} We investigated the effect of changes to discount and tremble rates. For our main analysis discount and tremble rates remained equal. The ratio of discount to tremble, however, can play an important role in model behavior \cite{skyrms2000dynamic}. Therefore we varied the ratio of discount to tremble and used 1 as a baseline. As this ratio is decreased (e.g. discount < tremble) the number of signaling groups formed also decreases. This is because when tremble is much larger than discount, it is hard for agents to form insulated subgroups. If tremble is large enough, this means most or all agents will converge to form a single signaling group. Conversely, when the ratio of discount to tremble increases (e.g. discount > tremble) the  number of signaling groups formed also increases. This occurs because now agents can more easily can form insulated groups. Higher discounting allows agents to more quickly forget undesired interactions, and low tremble prevents agents from mistakenly interacting outside of their group. Across a wide range of ratios, however, we find the continued formation of all three distinct signaling groups. 

When discount and tremble rates are set to 0, our results more drastically change. First network discount was set to 0, and strategy discount, network tremble, and strategy tremble remained 0.01. Under this setting, the population always converged to a single signaling group. This is because without discounting network weights remain more uniform, and the population is too well mixed to support separate groups (or hybrid solutions) forming within the seed. When strategy discount is set to 0, strategy weights remain more uniform and the signaling environment is too noisy for any subset of agents to learn an effective strategy. Alternatively when strategy tremble is set to 0, the accumulation of strategy weights may lock an agent into a sub-optimal strategy. This means semi-effective signaling groups can form in the population, but some agents become stuck on a strategy and cannot tremble into discovering the efficient signaling strategy for their group. Finally, when network tremble is set to 0, agents may become locked into a sub-optimal network relationship, but these agents can adapt their strategy to form efficient signaling groups given their pool of interactions. In this setting agents can still form distinct homogeneous and hybrid signaling groups within a seed. 

\paragraph{Symmetric network learning.}
Throughout our simulations network weights are updated asymmetrically, meaning only the agent who initiated the interaction (the visitor) updates their outgoing link weight based on the received payoff. If network learning is symmetric, meaning both agents in an interaction update their outgoing link weights, the network fractures into small groups. Using our network thresholding approach, the average number of groups per seed in our baseline model was 2.3, while it was 7.4 for the model with symmetric updating. Just as in our baseline model, the model with symmetric updating still supported the frequent formation of all 3 types of strategy groups.

\cleardoublepage
\printbibliography[title={Supplemental References}]

@article{li2017fundamental,
  title={The fundamental advantages of temporal networks},
  author={Li, Aming and Cornelius, Sean P and Liu, Y-Y and Wang, Long and Barab{\'a}si, A-L},
  journal={Science},
  volume={358},
  number={6366},
  pages={1042--1046},
  year={2017},
  publisher={American Association for the Advancement of Science}
}

@book{searcy2005,
  title={The Evolution of Animal Communication: Reliability and Deception in Signaling Systems},
  author={Searcy, W. A. and Nowicki, S.},
  year={2005},
  publisher={Princeton University Press}
}

@book{jms-harper2003,
  title={Animal Signals},
  author={Maynard Smith, J. and Harper, D.},
  year={2003},
  publisher={Oxford University Press}
}

@book{bradbury2011,
  title={Principles of Animal Communication},
  author={Bradbury, J. W. and Vehrencamp, S. L.},
  year={2011},
  publisher={Sinauer Associates}
}

@article{hu2011reinforcement,
  title={Reinforcement learning in signaling game},
  author={Hu, Yilei and Skyrms, Brian and Tarr{\`e}s, Pierre},
  journal={arXiv preprint arXiv:1103.5818},
  year={2011}
}

@article{nowak1999evolution,
  title={The evolution of language},
  author={Nowak, Martin A and Krakauer, David C},
  journal={Proceedings of the National Academy of Sciences},
  volume={96},
  number={14},
  pages={8028--8033},
  year={1999},
  publisher={National Acad Sciences}
}

@article{huttegger2014some,
  title={Some dynamics of signaling games},
  author={Huttegger, Simon and Skyrms, Brian and Tarres, Pierre and Wagner, Elliott},
  journal={Proceedings of the National Academy of Sciences},
  volume={111},
  number={Supplement 3},
  pages={10873--10880},
  year={2014},
  publisher={National Acad Sciences}
}

@article{skyrms2000dynamic,
	Author = {Skyrms, Brian and Pemantle, Robin},
	Journal = {Proceedings of the National Academy of Sciences},
	Number = {16},
	Pages = {9340--9346},
	Publisher = {National Acad Sciences},
	Title = {A dynamic model of social network formation},
	Volume = {97},
	Year = {2000}
}

@book{skyrms2010signals,
  title={Signals: Evolution, learning, and information},
  author={Skyrms, Brian},
  year={2010},
  publisher={OUP Oxford}
}

@article{huttegger2010evolutionary,
  title={Evolutionary dynamics of Lewis signaling games: signaling systems vs. partial pooling},
  author={Huttegger, Simon M and Skyrms, Brian and Smead, Rory and Zollman, Kevin JS},
  journal={Synthese},
  volume={172},
  number={1},
  pages={177--191},
  year={2010},
  publisher={Springer}
}

@book{lewis1969convention,
  title={Convention: A philosophical study},
  author={Lewis, D.},
  year={1969},
  publisher={Harvard University Press},
  city={Cambridge, MA, USA}
}

@book{pinker2003language,
  title={The language instinct: How the mind creates language},
  author={Pinker, Steven},
  year={2003},
  publisher={Penguin UK}
}

@article{lieberman2007quantifying,
  title={Quantifying the evolutionary dynamics of language},
  author={Lieberman, Erez and Michel, Jean-Baptiste and Jackson, Joe and Tang, Tina and Nowak, Martin A},
  journal={Nature},
  volume={449},
  number={7163},
  pages={713--716},
  year={2007},
  publisher={Nature Publishing Group}
}

@article{vespignani2012modelling,
  title={Modelling dynamical processes in complex socio-technical systems},
  author={Vespignani, Alessandro},
  journal={Nature Physics},
  volume={8},
  number={1},
  pages={32--39},
  year={2012},
  publisher={Nature Publishing Group}
}

@article{perc2010coevolutionary,
  title={Coevolutionary games—a mini review},
  author={Perc, Matja{\v{z}} and Szolnoki, Attila},
  journal={BioSystems},
  volume={99},
  number={2},
  pages={109--125},
  year={2010},
  publisher={Elsevier}
}

@article{foley2021avoiding,
  title={Avoiding the bullies: The resilience of cooperation among unequals},
  author={Foley, Michael and Smead, Rory and Forber, Patrick and Riedl, Christoph},
  journal={PLoS Computational Biology},
  volume={17},
  number={4},
  pages={e1008847},
  year={2021},
  publisher={Public Library of Science San Francisco, CA USA}
}

@article{newman2003structure,
  title={The structure and function of complex networks},
  author={Newman, Mark EJ},
  journal={SIAM Review},
  volume={45},
  number={2},
  pages={167--256},
  year={2003},
  publisher={SIAM}
}

@article{barabasi1999emergence,
  title={Emergence of scaling in random networks},
  author={Barab{\'a}si, Albert-L{\'a}szl{\'o} and Albert, R{\'e}ka},
  journal={Science},
  volume={286},
  number={5439},
  pages={509--512},
  year={1999},
  publisher={American Association for the Advancement of Science}
}

@inproceedings{leskovec2005graphs,
  title={Graphs over time: densification laws, shrinking diameters and possible explanations},
  author={Leskovec, Jure and Kleinberg, Jon and Faloutsos, Christos},
  booktitle={Proceedings of the Eleventh ACM SIGKDD International Conference on Knowledge Discovery in Data Mining},
  pages={177--187},
  year={2005}
}

@ARTICLE{Rand2014,
   author = {Rand, D. and Nowak, M. and Fowler, J. and Christakis, N.},
   title = {Static network structure can stabilize human cooperation},
   journal = {Proceedings of the National Academy of Sciences},
   volume = {111},
   pages = {17093--17098},
   year = {2014}
}

@article{SkyrmsPemantle2000,
    author = {Skyrms, B. and Pemantle, R.},
    title = {A dynamic model of social network formation},
    journal = {Proceedings of the National Academy of Sciences},
    volume = {97},
    number = {16},
    pages = {9340-9346},
    year = {2000}
}

@article{SinghEtAl2019,
  title={Anti-modular nature of partially bipartite networks makes them infra small-world},
  author={Singh, Aradhana and Ashraf, Md and Sinha, Sitabhra and others},
  journal={arXiv preprint arXiv:1902.02668},
  year={2019}
}

@article{PhysRevLett.87.198701,
  title = {Efficient Behavior of Small-World Networks},
  author = {Latora, Vito and Marchiori, Massimo},
  journal = {Phys. Rev. Lett.},
  volume = {87},
  issue = {19},
  pages = {198701},
  numpages = {4},
  year = {2001},
  month = {Oct},
  publisher = {American Physical Society}
}

@book{chaudhuri2009experiments,
  title={Experiments in economics: playing fair with money},
  author={Chaudhuri, Ananish},
  year={2009},
  publisher={Routledge},
  city={London, UK}
}

@article{Fulker2021,
    author = {Fulker, Z. and Forber, P. and Smead, R. and Riedl, C.},
    title = {Spite is contagious in dynamic networks},
    journal = {Nat Commun},
    volume = {12},
    number = {260},
    year = {2021}
}

@book{skyrms2004stag,
  title={The stag hunt and the evolution of social structure},
  author={Skyrms, Brian},
  year={2004},
  publisher={Cambridge University Press}
}

@article{FoleyEtAl2018,
    author = {Michael Foley  and Patrick Forber  and Rory Smead  and Christoph Riedl },
    title = {Conflict and convention in dynamic networks},
    journal = {Journal of the Royal Society Interface},
    volume = {15},
    number = {140},
    pages = {20170835},
    year = {2018}
}

@article{erev1995,
	Author = {Roth, A. and Erev, I.},
	Journal = {Games and Economic Behavior},
	Pages = {164--212},
	volume = {8},
	Title = {Learning in extensive-form games: experimental
    data and simple dynamic models in the
    intermediate term},
	Year = {1995}
}

@article{erev1998predicting,
	Author = {Erev, Ido and Roth, Alvin E},
	Journal = {American Economic Review},
	Pages = {848--881},
	Publisher = {JSTOR},
	Title = {Predicting how people play games: reinforcement learning in experimental games with unique, mixed strategy equilibria},
	Year = {1998}
}

@article{Gershman2017,
	Author = {Gershman, Samuel and Daw, Nathaniel},
	Journal = {Annual Review of Psychology},
	Pages = {101--128},
	volume = {68},
	Title = {Reinforcement learning and episodic memory in humans and animals: an integrative framework},
	Year = {2017}
}

@ARTICLE{Pacheco2006,
   author = {Pacheco, J. and Traulsen, A. and Nowak, M.},
   title = {Coevolution of strategy and structure in complex networks with dynamical linking},
   journal = {Physical Review Letters},
   volume = {97},
   pages = {258103},
   year = {2006}
}

@ARTICLE{Rand2011,
   author = {Rand, D. and Arbesman, S. and Christakis, N.},
   title = {Dynamic social networks promote cooperation in experiments with humans},
   journal = {Proceedings of the National Academy of Sciences},
   volume = {108},
   pages = {19193--19198},
   year = {2011}
}

@ARTICLE{Goyal2005,
   author = {Goyal, S. and Vega-Redondo, F.},
   title = {Network formation and social coordination},
   journal = {Games and Economic Behavior},
   volume = {50},
   pages = {178--207},
   year = {2005}
}

@ARTICLE{Granovetter1973,
   author = {Granovetter, M.},
   title = {The strength of weak ties},
   journal = {American Journal of Sociology},
   volume = {78},
   pages = {1360--1380},
   year = {1973}
}

@ARTICLE{Barrat2004,
   author = {Barrat, A. and Barthelemy, M. and Pastor-Satorras, R. and Vespignani, A.},
   title = {The architecture of complex weighted networks},
   journal = {Proceedings of the National Academy of Sciences},
   volume = {101},
   pages = {3747--3752},
   year = {2004}
}

@ARTICLE{Yook2001,
   author = {Yook, S. and Jeong, H. and Barabasi, A. and Tu, Y.},
   title = {Weighted evolving networks.},
   journal = {Physical Review Letters},
   volume = {86},
   pages = {5835--5838},
   year = {2001}
}

@ARTICLE{Castellano2009,
   author = {Castellano, C. and Fortunato, S. and Loreto, V.},
   title = {Statistical physics of social dynamics},
   journal = {Rev. Mod. Phys.},
   volume = {81},
   pages = {591},
   year = {2009}
}

@ARTICLE{Ke2008,
   author = {Ke, J.  and Gong, T. and William, S.},
   title = {Language change in social networks},
   journal = {Commun. Comput. Phys},
   volume = {3},
   number = {4},
   pages = {935-949},
   year = {2008}
}

@ARTICLE{Huford1989,
   author = {Hurford, J.},
   title = {Biological evolution of the Saussurean sign as a component of the language acquisition device},
   journal = {Lingua},
   volume = {77},
   number = {2},
   pages = {187-222},
   year = {1989}
}

@book{Chomsky1965,
  title={Aspects of the theory of syntax},
  author={Chomsky, Noam},
  year={1965},
  publisher={MIT Press},
  city={Cambridge, MA}
}

@ARTICLE{Pinker1990,
   author = {Pinker, S. and Bloom, P.},
   title = {Natural language and natural selection},
   journal = {Behavioral and Brain Sciences},
   volume = {13},
   number = {4},
   pages = {707-784},
   year = {1990}
}

@book{skyrms2014evolution,
  title={Evolution of the social contract},
  author={Skyrms, Brian},
  year={2014},
  publisher={Cambridge University Press}
}

@book{Vespignani2008DynProcesses,
  title={Dynamical Processes on Complex Networks},
  author={Barrat, Alain. and Barth{\'e}lemy, Marc. and Vespignani, Alessandro.},
  year={2008},
  publisher={Cambridge University Press}
}

@ARTICLE{Nowak1999Bio,
   author = {Nowak, M. and Plotkin, J. and Krakauer, D.},
   title = {The Evolutionary Language Game},
   journal = {J. Theor. Biol},
   volume = {200},
   number = {2},
   pages = {147-162},
   year = {1999}
}

@ARTICLE{Baronchelli2006,
   author = {Baronchelli, A. and Felici, M. and Loreto, V. and Caglioti, E. and Steels, L.},
   title = {Sharp transition towards shared vocabularies in multi-agent systems},
   journal = {J. Stat. Mech. Theor. and Experiment},
   volume = {2006},
   pages = {P06014},
   year = {2006}
}

@ARTICLE{Baronchelli2008,
   author = {Baronchelli, A. and Loreto, V. and Steels, L.},
   title = {In-depth analysis of the Naming Game dynamics: the homogeneous mixing case},
   journal = {Intl. J. of Modern Physics C},
   volume = {19},
   number = {5},
   pages = {785-812},
   year = {2008}
}

@ARTICLE{Steels2005,
   author = {Steels, L. and Belpaeme T.},
   title = {Coordinating perceptually grounded categories through language: a case study for colour.},
   journal = {Behav. Brain. Sci.},
   volume = {28},
   number = {4},
   pages = {489-529},
   year = {2005}
}

@ARTICLE{Baronchelli2010,
   author = {Baronchelli, A. and Gong, T. and Puglisi, A. and Loreto, V.},
   title = {Modeling the emergence of universality in color naming patterns},
   journal = {Proceedings of the National Academy of Sciences},
   volume = {107},
   number = {6},
   pages = {2403-2407},
   year = {2010}
}

@ARTICLE{LaCroix2018,
   author = {LaCroix, T.},
   title = {On salience and signaling in sender receiver games: Partial pooling, learning, and focal points},
   journal = {Synthese},
   volume = {197},
   pages = {1725–1747},
   year = {2018}
}

@ARTICLE{LaCroix2020,
   author = {LaCroix, T.},
   title = {Evolutionary Explanations of Simple Communication: Signalling Games and Their Models},
   journal = {Journal for General Philosophy of Science},
   volume = {51},
   pages = {19–43},
   year = {2020}
}

@ARTICLE{Barrett2006,
   author = {Barrett, J},
   title = {Numerical Simulations of the Lewis Signaling Game: Learning Strategies, Pooling Equilibria, and the Evolution of Grammar},
   journal = {UC Irvine: Institute for Mathematical Behavioral Sciences},
   year = {2006}
}

@ARTICLE{Zollman2005,
   author = {Zollman, K. J. S.},
   title = {Talking to neighbors: The evolution of regional meaning},
   journal = {Philosophy of Science},
   volume = {72},
   number = {1},
   pages = {69–85},
   year = {2005}
}

@ARTICLE{Wagner2009,
   author = {Wagner, E. O.},
   title = {Communication and structured correlation},
   journal = {Erkenntnis},
   volume = {71},
   number = {3},
   pages = {377–393},
   year = {2009}
}

@article{muhlenbernd2011learning,
  title={Learning with neighbours},
  author={M{\"u}hlenbernd, Roland},
  journal={Synthese},
  volume={183},
  number={1},
  pages={87--109},
  year={2011},
  publisher={Springer}
}

@article{hofbauer2008feasibility,
  title={Feasibility of communication in binary signaling games},
  author={Hofbauer, Josef and Huttegger, Simon M},
  journal={Journal of Theoretical Biology},
  volume={254},
  number={4},
  pages={843--849},
  year={2008},
  publisher={Elsevier}
}

@article{zollman2005talking,
  title={Talking to neighbors: The evolution of regional meaning},
  author={Zollman, Kevin JS},
  journal={Philosophy of Science},
  volume={72},
  number={1},
  pages={69--85},
  year={2005},
  publisher={The University of Chicago Press}
}

@article{barrett2009role,
  title={The role of forgetting in the evolution and learning of language},
  author={Barrett, Jeffrey and Zollman, Kevin JS},
  journal={Journal of Experimental \& Theoretical Artificial Intelligence},
  volume={21},
  number={4},
  pages={293--309},
  year={2009},
  publisher={Taylor \& Francis}
}

@article{argiento2009learning,
  title={Learning to signal: Analysis of a micro-level reinforcement model},
  author={Argiento, Raffaele and Pemantle, Robin and Skyrms, Brian and Volkov, Stanislav},
  journal={Stochastic Processes and their Applications},
  volume={119},
  number={2},
  pages={373--390},
  year={2009},
  publisher={Elsevier}
}

@article{franke2020vagueness,
  title={Vagueness and imprecise imitation in signalling games},
  author={Franke, Michael and Correia, Jos{\'e} Pedro},
  journal={British Journal for the Philosophy of Science},
  year={2020},
  publisher={The University of Chicago Press}
}

@incollection{muhlenbernd2012signaling,
  title={Signaling conventions: Who learns what where and when in a social network?},
  author={M{\"u}hlenbernd, Roland and Franke, Michael},
  booktitle={The evolution of language},
  pages={242--249},
  year={2012},
  publisher={World Scientific}
}

@article{pawlowitsch2007finite,
  title={Finite populations choose an optimal language},
  author={Pawlowitsch, Christina},
  journal={Journal of Theoretical Biology},
  volume={249},
  number={3},
  pages={606--616},
  year={2007},
  publisher={Elsevier}
}

@article{beggs2005convergence,
  title={On the convergence of reinforcement learning},
  author={Beggs, Alan W},
  journal={Journal of economic theory},
  volume={122},
  number={1},
  pages={1--36},
  year={2005},
  publisher={Elsevier}
}

@article{tomassini2010,
  title={Coordination games on dynamical networks},
  author={Tomassini M, Pestelacci E. },
  journal={Games},
  volume={3},
  number={1},
  pages={242-261},
  year={2010},
  publisher={MDPI}
}

@article{zimmermann2004coevolution,
  title={Coevolution of dynamical states and interactions in dynamic networks},
  author={Zimmermann, Mart{\'\i}n G and Egu{\'\i}luz, V{\'\i}ctor M and San Miguel, Maxi},
  journal={Physical Review E},
  volume={69},
  number={6},
  pages={065102},
  year={2004},
  publisher={APS}
}

@article{godfrey2013communication,
  title={Communication and common interest},
  author={Godfrey-Smith, Peter and Mart{\'\i}nez, Manolo},
  journal={PLoS computational biology},
  volume={9},
  number={11},
  pages={e1003282},
  year={2013},
  publisher={Public Library of Science San Francisco, USA}
}

@article{huttegger2010dynamic,
  title={Dynamic stability and basins of attraction in the Sir Philip Sidney game},
  author={Huttegger, Simon M and Zollman, Kevin JS},
  journal={Proceedings of the Royal Society B: Biological Sciences},
  volume={277},
  number={1689},
  pages={1915--1922},
  year={2010},
  publisher={The Royal Society}
}
\end{refsection}

\end{document}